\documentclass[reprint,aps,pra,superscriptaddress,longbibliography]{revtex4-2}
\usepackage{physics,mathtools,amsmath,amssymb}
\usepackage{xfrac}
\usepackage{graphicx}

\usepackage{microtype}
\usepackage{balance}          
\usepackage{titlesec}         

\usepackage{ifthen}
\usepackage{caption}
\usepackage{subfig}
\usepackage{natbib}
\usepackage[colorlinks,linkcolor=blue,citecolor=blue, urlcolor  = blue]{hyperref}
\usepackage{float}
\graphicspath{{images/}}

\begin{document}

\title{Properties of a Three-Level $\Lambda$-Type Atom Driven by Coherent and Stochastic Fields}%
\author{Sajad Ahmadi}
\affiliation{Department of Physics, Institute for Advanced Studies in Basic Sciences (IASBS), 45137-66731, Zanjan, Iran}

\author{Mohsen Akbari}
\email[Corresponding author email:]{mohsen.akbari@khu.ac.ir}
\affiliation{Quantum Optics Laboratory, Department of Physics, Kharazmi  University, 16315-1355 Tehran, Iran}

\author{Shahpoor Saeidian}
\affiliation{Department of Physics, Institute for Advanced Studies in Basic Sciences (IASBS), 45137-66731, Zanjan, Iran}

\author{Ali Motazedifard}
\affiliation{Department of Physics, College of Science, University of Tehran, Tehran, 14395-547,Iran}
\affiliation{Quantum Sensing Lab, Quantum Metrology Group, Iranian Center for Quantum Technologies (ICQT), Tehran, 15998-14713,Iran}
\date{\today}%

\begin{abstract}
We present a theoretical investigation of a three-level $\Lambda$-type atom driven by a strong coherent laser and a weak stochastic field exhibiting amplitude and phase fluctuations. The stochastic field is modeled as a complex Gaussian–Markovian random process with finite bandwidth to describe realistic laser noise. Using the Born–Markov and rotating-wave approximations, we derive a Lindblad-form master equation that incorporates spontaneous emission and noise-induced terms, and we solve for the steady-state regime. We examine level populations in both the bare and dressed bases and compute the incoherent resonance-fluorescence spectrum. Our analysis shows that the stochastic drive is not merely a source of decoherence but a versatile control parameter. By detuning the stochastic-field central frequency relative to the coherent drive (especially for narrow bandwidths), we observe pronounced changes in emission characteristics, including selective enhancement or suppression, and reshaping of the multi-peaked fluorescence spectrum when the detuning matches the generalized Rabi frequency. Numerical results reveal nontrivial steady-state modifications distinct from purely coherent driving, enabling precise control of populations and suggesting applications in quantum control, quantum technologies, spectroscopy, and noise-assisted manipulation of atomic systems.
\end{abstract}

\maketitle

\section{Introduction}
The interaction between light and matter, particularly in the presence of field fluctuations, has been a central focus of research for several decades. In realistic experimental systems, it is virtually impossible to produce an electromagnetic driving field completely free of amplitude or phase noise; consequently, many theoretical and experimental studies have examined the atomic response to driving fields exhibiting amplitude or phase fluctuations. Foundational work from the 1970–2002 established analytical noise models (e.g., phase-diffusion and chaotic-field descriptions) and examined their effects on level populations, resonance fluorescence, and coherence properties (see, e.g., \cite{Vemuri,Dalton1982,Freedhoff,Zhang2002,Georges,Vogel1983,PhysRevA.68.033809,Li2001,Li2002}). More recent studies (2010–2024) have extended these approaches using Gaussian–Markovian and bandwidth-limited stochastic models, numerical simulations, and applications to cold atoms and quantum technologies (for more details see \cite{Santos2021,Sen2023,Yaping}).\\

The most common models, which include pump-field fluctuations, have therefore been used in the literature. \cite{Vemuri,Georges,Zhang2002}. Below we summarize the most common models:\\
1. Phase-diffusion model: Only phase fluctuations are included; this model corresponds to a single-mode laser with a steady intensity and admits well-known analytical solutions \cite{Vemuri,Georges}.\\
2. Chaotic field model: The complex field amplitude is treated as a two-dimensional Gaussian process, so both amplitude and phase fluctuate; this models a multimode laser with many uncorrelated modes \cite{Vemuri,Georges,Zhang2002}.\\
3. Gaussian-amplitude model: Only amplitude fluctuations are considered; amplitude noise in this model can be stronger than in the chaotic-field model \cite{Georges,Zhang2002}.\\
4. Jump models: Amplitude, phase, or frequency evolve via discontinuous (Markovian) jump processes \cite{Vemuri}.\\
5. Gaussian–Markovian model: Fluctuations are simultaneously Gaussian and Markovian; this model satisfies both Gaussian statistics and memoryless dynamics and is particularly convenient for open-system treatments using the Lindblad master equation \cite{Lamon,Rasmussen}.

Despite this progress, to the best of our knowledge, no studies have addressed three-level $\Lambda$-type atoms driven by classical fields that exhibit both amplitude and phase fluctuations. This gap is important because  $\Lambda$-type systems underpin applications such as Coherent Population Trapping (CPT) clocks \cite{Ahmadi,Vanier2005,Guo2009} and quantum memories based on Electromagnetically Induced Transparency (EIT) \cite{Ahmadi,Fleischhauer}.\\ 
In many laboratories, light sources present finite-bandwidth amplitude and phase fluctuations that cannot be neglected in practice. Here we show that, when modeled as a complex Gaussian–Markovian drive, such stochastic features can be repurposed as inexpensive, readily accessible control parameters for  $\Lambda$-type atoms: by tuning noise bandwidth and central frequency one can manipulate dressed-state occupations and incoherent fluorescence, with direct implications for robust state preparation, error mitigation, and noise-aware protocols in quantum information and sensing.

Many noisy optical environments display approximately Gaussian statistics and short correlation times, making the Gaussian–Markovian description a natural and practical choice. When the environment correlation time ($\tau_S$) is much shorter than the characteristic timescale of the quantum system, the Markov approximation is well justified and enables an effective master-equation description of the system dynamics.

In this work, we model the applied stochastic field as a Gaussian–Markovian process and incorporate it into a Lindblad-type master equation. This approach captures the essential statistical features of realistic pump noise while remaining tractable for analytical insight and numerical simulation.

 We consider a three-level atom in the $\Lambda$ configuration driven simultaneously by two fields: a strong coherent field and a weak finite-bandwidth stochastic field whose amplitude and phase fluctuate. We study how the stochastic field affects key observables—in particular the steady-state populations and the resonance-fluorescence spectrum—and we identify conditions under which pump-noise engineering can be used to control atomic radiative properties. We show that tuning the central frequency of the finite-bandwidth stochastic field provides an efficient means to control the atomic populations and to reshape the resonance-fluorescence spectrum.\\

The remainder of the paper is organized as follows. In Sec.\ref{sec:model} we introduce the three-level $\Lambda$-type atomic model, the coherent driving fields, and the stochastic field used to model environmental fluctuations. In Sec.\ref{sec:master} we derive the master equation governing the system dynamics and discuss the approximations and parameter definitions employed. In Sec.\ref{sec:steady} we analyze the steady-state population distribution using both bare and dressed-state descriptions to elucidate the interplay between coherent and stochastic driving. In Sec.\ref{sec:fluorescence} we devote our attention to the incoherent resonance fluorescence spectrum: we derive analytical expressions and present numerical results that highlight noise-induced modifications of spectral features. Finally, in Sec.\ref{sec:conclusion}, we summarize our conclusions.
\section{The system Model\label{sec:model}}
The Gaussian–Markovian model can be incorporated into the Lindblad master-equation framework, providing an effective description of a system coupled to a noisy reservoir. In quantum optics it is particularly well suited for treating pump-field fluctuations in light scattering from atoms and provides a convenient basis for analytical and numerical calculations.\\

We consider the total system (atom + environment) subject to an external driving field that comprises a coherent component and a noisy (stochastic) component; the noisy component effectively acts as a reservoir or environment for the atom. The applied field is taken as
\begin{align}
	\mathbf{E}= \mathbf{\hat{e}}\Big(E_c e^{-i\omega_L t} + E_s (t) e^{-i\omega_s t}\Big) +c.c.
	\label{b1}
\end{align}
where $\mathbf{\hat{e}}$ is the unit polarization vector and $E_c$ is the coherent part of the field envelope. The phase and amplitude fluctuations are described by $E_s(t) = \xi(t) e^{-i\varphi(t)}$, where \( \xi(t) \) is the real amplitude and \( \varphi(t) \) is the phase of the field \cite{Georges}.\\
 It is assumed that $E_s(t)$ is a zero-mean complex Gaussian–Markovian random process with non-vanishing correlation functions \cite{Vemuri,meystre,PhysRevLett.37.1383,orszag}.  The Gaussian–Markovian model yields, for $t \neq t'$,
\begin{align}
	\langle E_s (t) E_s ^* (t')\rangle =D'\kappa e^{-\kappa \abs{t-t'}}, \ \  \langle E_s(t) E_s  (t') \rangle =0 
	\label{b2}
\end{align}
with $D'$ denoting the strength of the stochastic process and $\kappa$ its bandwidth. Equation \eqref{b2} describes a classical field with fluctuations in both amplitude and phase \cite{Zhou1998}.

\begin{figure}[H]
	\centering
	\includegraphics[width=0.55\linewidth]{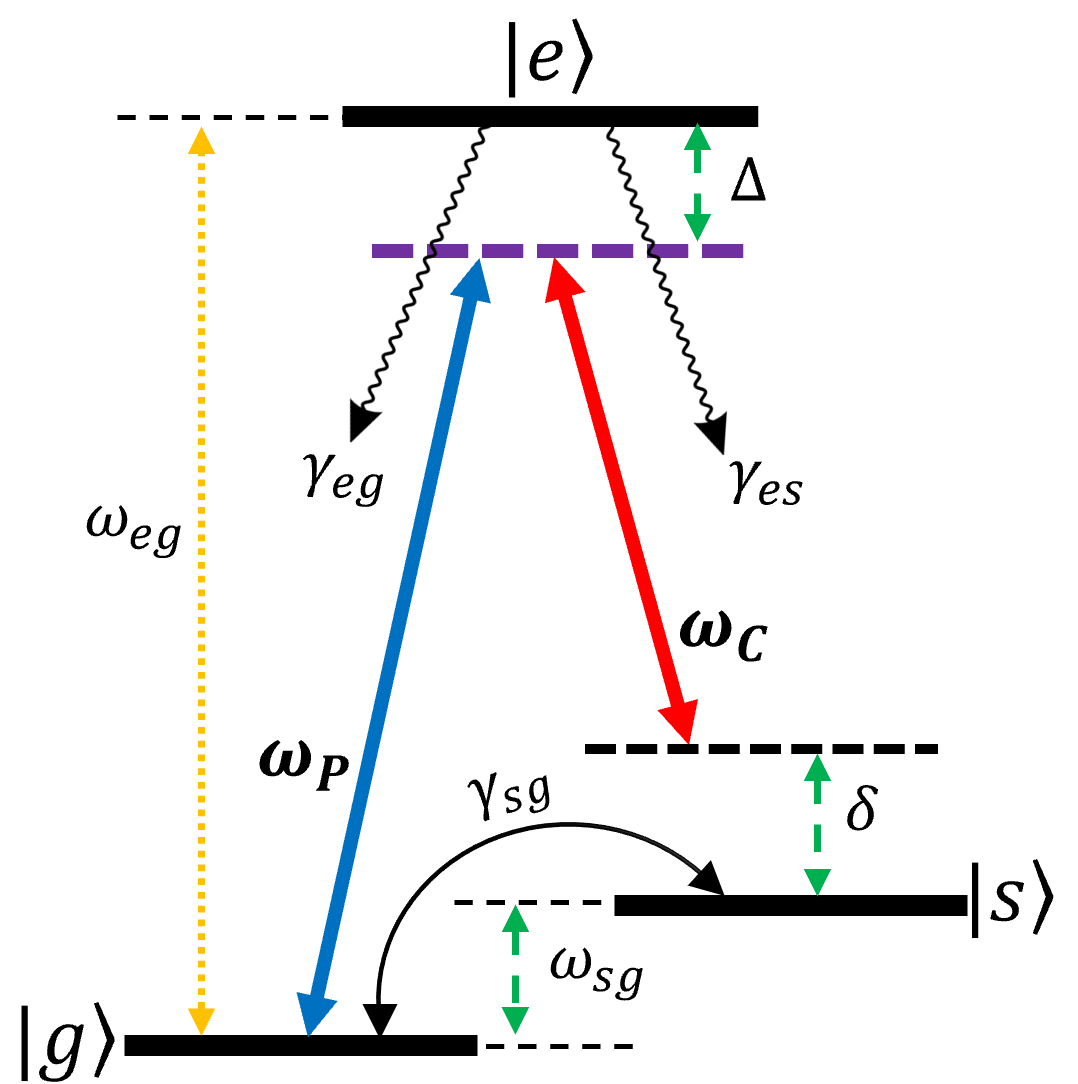}
	\caption{Schematic of the \(\Lambda\)-type atomic system. The transitions \(\ket{g}\leftrightarrow\ket{e}\) and \(\ket{s}\leftrightarrow\ket{e}\) are driven by fields with frequencies \(\omega_P\) and \(\omega_C\), respectively. The system is characterized by one-photon detunings \(\Delta\) and \(\Delta'\), two-photon detuning \(\delta\), and decay/decoherence rates \(\gamma_{eg}\), \(\gamma_{es}\), and \(\gamma_{sg}\). Transition frequencies are defined as \(\omega_{ij} = \omega_i - \omega_j\) for \(i,j \in \{e,g,s\}\).}
	\label{fig:lmbd}
\end{figure}

Accordingly, rather than applying a purely monochromatic laser, we consider a driving field composed of coherent and stochastic parts. This choice improves experimental realism, since perfectly noise-free coherent fields are not available. We focus on the regime where the atom is driven by a strong coherent field and a weak stochastic field of arbitrary bandwidth. Because the stochastic component acts as noise on the coherent drive, the coherent intensity is assumed to be much larger than the stochastic intensity.\\
Furthermore, the stochastic-field bandwidth \(\kappa\) is assumed much larger than the atomic linewidth \(\gamma\).  The atomic relaxation time and the reservoir correlation time are then approximately \(\tau_A\simeq 1/\gamma\) and \(\tau_S\simeq 1/\kappa\), respectively. Hence \(\kappa\gg\gamma\) implies \(\tau_S\ll\tau_A\): the reservoir decorrelates on a timescale much shorter than the atomic relaxation time.  This separation of timescales justifies the Born–Markov approximation and the resulting effective master-equation description of the atomic dynamics.
The schematic of a $\Lambda$-type atomic configuration used in this study is depicted in Fig.\ref{fig:lmbd} and comprises an excited state \(\ket{e}\), a ground state \(\ket{g}\), and a metastable state \(\ket{s}\). The system is driven by a classical field with two components: (i) a strong coherent field with frequencies \(\omega_P\) and \(\omega_C\) driving the \(\ket{g}\leftrightarrow\ket{e}\) and \(\ket{s}\leftrightarrow\ket{e}\) transitions, respectively, and (ii) a stochastic field with central frequency \(\omega_s\) and complex, fluctuating amplitude \(E_s(t)\), treated semiclassically.\\
These fields induce transitions with associated spontaneous emission from \(\ket{e}\) to \(\ket{g}\) and \(\ket{s}\) at rates \(\gamma_{eg}\) and \(\gamma_{es}\), respectively, and decoherence between \(\ket{g}\) and \(\ket{s}\) at rate \(\gamma_{sg}\). The detunings are defined as
\begin{equation}
	 \begin{aligned} 
	&\Delta=\omega_{eg}-\omega_{P} \ \ \  , \ \ \omega_{eg}=\omega_e -\omega_g \\ 
	&\Delta'=\omega_{es}-\omega_{C}  \ \ , \ \ \omega_{es}=\omega_e -\omega_s \\
	& \delta = \Delta -\Delta' = \omega_{eg} - \omega_{es}
   \end{aligned}
	\label{x4}
\end{equation}
For our analysis, we assume the coherent field drives both transitions with equal strength ($\omega_P = \omega_C = \omega_L$) and that the dipole moments are equal ($|\mu_{eg}| = |\mu_{es}|$). As a result, the Rabi frequencies are given by ($\Omega_P = \Omega_C \equiv \Omega$), and the system is set at two-photon resonance ($\delta = 0$).
\section{Master equation\label{sec:master}}
To analyze the system dynamics, we use the master equation for the full system (atom + environment). In a frame rotating at frequency \(\omega_L\), and with \(\hbar=1\), the equation takes the form
\begin{equation}
\dot{\rho} = -i[H,\rho] + \mathcal{L}_A\rho,
	\label{x1}
\end{equation}
where $H$ is the total Hamiltonian in the interaction picture with the rotating-wave approximation (RWA) applied, and $\mathcal{L}_A$ is the Liouvillian superoperator describing atomic dissipation. We decompose
\begin{equation}
	H = H_0 + H_{ac} + H_{as},
	\label{x2}
\end{equation}
where $H_0$ is the free atomic Hamiltonian, $H_{ac}$ the interaction with the coherent field, and $H_{as}$ the interaction with the stochastic field. The explicit form of $ H_0 + H_{ac} $ is defined by $ H_{\Lambda} $ \cite{Ahmadi,Bergmann_2019,Shore2017,vit}
\begin{equation}
	H_{\Lambda}=\Delta\sigma_{ee}+\Omega \Big(\sigma_{eg}+\sigma_{es}+h.c.\Big)
	\label{x10}
\end{equation}
where $\sigma_{ij}=\ketbra{i}{j}$ with ($i,j\in{g,e,s}$) are the atomic transition operators. The state vectors are represented in the column form $ \ket{g}=(1\ 0\ 0)^T,  \ket{e}=(0\ 1\ 0)^T$, and $\ket{s}=(0\ 0\ 1)^T$. 
The remaining term in Eq.\eqref{x2}, $H_{as}$, describes the atom's coupling to the fluctuating field. A convenient form is
 \begin{equation}
  \begin{aligned}
  		H_{as}=&\dfrac{1}{2}\Big[\Big(\mathcal{F}_g(t)\sigma_{eg}+\mathcal{F}_s(t)\sigma_{es}\Big)e^{-i\eta t} \\
  		& + \Big(\mathcal{F}^*_g(t)\sigma_{ge}+ \mathcal{F}^*_s(t)\sigma_{se}\Big)e^{i\eta t} \Big]
  \end{aligned}
	\label{x18}
\end{equation}
where $ \eta=\omega_s - \omega_L $ and $	\mathcal{F}_i(t)$ is amplitude of stochastic-field and defined as
\begin{equation}
	\mathcal{F}_g(t)=-\bra{e}2\boldsymbol{\mu.}\hat{\mathbf{e}}E_s\ket{g} ,\\ \mathcal{F}_s(t)=-\bra{e}2\boldsymbol{\mu.}\hat{\mathbf{e}}E_s\ket{s}
	\label{x13}
\end{equation}
Since $E_s $ is a complex Gaussian–Markovian process, in terms of Eq.\eqref{b2}, we can write for Eq.\eqref{x13} 
\begin{equation}
	\begin{aligned} 
		&\langle \mathcal{F}_i(t)\mathcal{F}_j(t')\rangle=0 \ \ \ , \ \ \ t\neq t' \\[8pt]
		&\langle \mathcal{F}_i(t)\mathcal{F}^*_j(t')\rangle=D\kappa e^{-\kappa \vert t-t' \vert}  , \ \ \ i,j=g,s
	\end{aligned}
	\label{x14}
\end{equation}
where $D$ is the strength of the stochastic field and $\kappa$ its bandwidth. \\
The term $\mathcal{L}_A \rho$ refers to the states of the atom affected by perturbations or environmental noise. Thus, it will only act on the reduced density operator of the system, which in this case is the atomic density operator. Consequently,
\begin{equation}
\begin{aligned}
	 \mathcal{L}_A \rho_A=&\gamma \Big[ (\sigma_{ge}\rho_A\sigma_{eg}+\sigma_{se}\rho_A\sigma_{es})- \lbrace\sigma_{ee},\rho_A\rbrace \Big] \\
	 &+ \gamma_{sg} \Big[ \sigma_{gs}\rho_A\sigma_{sg} - \frac{1}{2}\lbrace\sigma_{ss},\rho_A\rbrace \Big]
\end{aligned}
	\label{x23}
\end{equation}
To eliminate the stochastic amplitudes, we apply the canonical transformation \(\tilde{\rho}(t)=e^{iH_{\Lambda}t}\rho e^{-iH_{\Lambda}t}\). Differentiating this and substituting Eq.\eqref{x1} yields the von Neumann equation in this picture. The damping term \(\mathcal{L}_A\rho\) is unaffected by this transformation, so we omit it temporarily  and reintroduce it later \cite{Zhou1998, Cirac}.
\begin{equation}
	\dot{\tilde{\rho}}(t) = -i[\tilde{H}_{as}(t),\tilde{\rho}(t)]
	\label{x30}
\end{equation}
After the Born-Markov approximation and returning to the original picture, the final master equation for the reduced atomic density matrix is
\begin{equation}
	\begin{aligned}
       \dot{\rho}_A=& -i[H_{\Lambda},\rho_A] + \mathcal{L}_A \rho_A\\
       &-\dfrac{1}{4}\!\Big(\Big[\sigma_{eg}\!+\!\sigma_{es}, [Z_- \!,\! \rho_A]\Big]+\Big[\sigma_{ge}\!+\!\sigma_{se}, [Z_+\! ,\! \rho_A]\Big]\Big) 
		\label{x51}
	\end{aligned}
\end{equation}
In the master equation, the effects of amplitude and phase fluctuations enter through the operators $Z_{\pm}$, with $Z_+=Z_-^{\dagger}$.
\begin{equation}
	Z_-  =
	\begin{pmatrix}
		\mathcal{M} & \mathcal{H} & \mathcal{M}\\
		\mathcal{N} & -2\mathcal{M} & \mathcal{N}\\
		\mathcal{M} & \mathcal{H} & \mathcal{M}\\
	\end{pmatrix}
	\label{x50}
\end{equation}
Here,
\begin{equation}
	\begin{aligned}
		&\mathcal{M}= \dfrac{\Omega}{2 R^2}  \Big[(\Delta +R)f(-1)-2 \Delta  f(0) + (\Delta -R)f(1)\Big]\\[9pt]
		&\mathcal{N}= \dfrac{2 \Omega ^2 }{R^2}\Big[-f(-1)+2 f(0)-f(1) \Big]\\[8pt]
		& \mathcal{H}= \dfrac{1}{4 R^2}\Big[(R +\Delta)^2f(-1) + 16\Omega^2f(0) + (R -\Delta)^2f(1) \Big]\\[9pt]
		&f(n)=\dfrac{D\ \kappa}{\kappa + i(\eta+ nR)}, \ \ n=0,\pm1 \\[9pt]
		&R=\sqrt{\Delta^2 + 8\Omega^2}
	\end{aligned}
	\label{x65}
\end{equation}
where \(R\) is the generalized Rabi frequency. \\
Eq.\eqref{x51} is most conveniently written in compact form
\begin{equation}
	\dot{\rho_A}=Q \rho_A + B
	\label{x71}
\end{equation}
Using the trace condition $\rho_{gg} + \rho_{ee} + \rho_{ss} = 1$, we eliminate $\rho_{ss}$. The quantity $\rho_A$ denotes the eight-component column vector formed from the remaining independent elements of the reduced atomic density matrix, $Q$ is the corresponding $8\times8$ evolution matrix, and $B$ is the eight-component column vector of constant coefficients. The explicit forms of $\rho_A$, $Q$, and $B$ are provided below.
\begin{align}
	&{\rho}_A =\left[{\rho}_{gg},\; {\rho}_{ge},\; {\rho}_{gs},\; {\rho}_{eg},\; {\rho}_{ee},\; {\rho}_{es},\; {\rho}_{sg},\; {\rho}_{se} \right]^T 	\label{xx70} \\[10pt]
	&B =\scalebox{0.8}{$ \! \left[\gamma _{sg} ,\! \dfrac{\mathcal{M}^{*}}{4},\! \! -\dfrac{\mathcal{H}^{*}}{4} ,\!   \dfrac{\mathcal{M}}{4},\! \dfrac{\Re(\mathcal{H})}{2} ,\!  -\dfrac{\mathcal{M}}{4}\! -\! i\Omega ,\!  
		-\dfrac{\mathcal{H}}{4},\!   -\dfrac{\mathcal{M}^{*}}{4}\! +\! i\Omega\right]^T $}
	\label{x70}
\end{align}
where $T$ denotes the transpose. For the matrix $Q$, we also write
\begin{widetext}
\begingroup
\setlength{\arraycolsep}{1.9pt} 
\renewcommand{\arraystretch}{2.5} 
\begin{equation}
	\resizebox{\textwidth}{!}{
		$\displaystyle
		\begin{aligned}
			&Q=\\
			&\begin{pmatrix}
				-\gamma _{sg}-\dfrac{\Re(\mathcal{H})}{2} & \dfrac{3\mathcal{M}}{4}+i\Omega & -\dfrac{\mathcal{H}}{4} &
				\dfrac{3\mathcal{M}^*}{4}-i\Omega &
				\gamma-\gamma _{sg}+\dfrac{\Re(\mathcal{H})}{2} &
				\dfrac{\mathcal{M}^*}{4} & -\dfrac{\mathcal{H}^*}{4} & \dfrac{\mathcal{M}}{4} \\[5pt]
				
				-\dfrac{\mathcal{M}^*}{2}+i\Omega &
				(-\gamma + i\Delta - \dfrac{3\mathcal{H}^*}{4}) &
				-\dfrac{\mathcal{M}^*}{4}+i\Omega &
				\dfrac{\mathcal{N}^*}{2} &
				-\dfrac{\mathcal{M}^*}{4}-i\Omega &
				\dfrac{\mathcal{N}^*}{2} &
				\dfrac{\mathcal{M}^*}{4} & -\dfrac{\mathcal{H}^*}{4} \\[5pt]
				
				-\dfrac{i}{2}\Im(\mathcal{H}) & \dfrac{3\mathcal{M}}{4}+i\Omega & \dfrac{-\gamma _{sg}-\Re(\mathcal{H})}{2} &
				\dfrac{\mathcal{M}^*}{4} & \dfrac{\big(\mathcal{H}+2\mathcal{H}^*\big)}{4} &
				\dfrac{3\mathcal{M}^*}{4}-i\Omega & 0 & \dfrac{\mathcal{M}}{4} \\[5pt]
				
				-\dfrac{\mathcal{M}}{2}-i\Omega & \dfrac{\mathcal{N}}{2} & \dfrac{\mathcal{M}}{4} &
				-\gamma - i\Delta - \dfrac{3\mathcal{H}}{4} & -\dfrac{\mathcal{M}}{4}+i\Omega &
				-\dfrac{\mathcal{H}}{4} & -\dfrac{\mathcal{M}}{4}-i\Omega & \dfrac{\mathcal{N}}{2} \\[5pt]
				
				0 & -\mathcal{M}-i\Omega & \dfrac{\Re(\mathcal{H})}{2} &
				-\mathcal{M}^*+i\Omega & -2\gamma - \dfrac{3\Re(\mathcal{H})}{2} &
				-\mathcal{M}^*+i\Omega & \dfrac{\Re(\mathcal{H})}{2} & -\mathcal{M}-i\Omega \\[5pt]
				
				\dfrac{\big(\mathcal{M}+2i\Omega\big)}{2} & \dfrac{\mathcal{N}}{2} &
				-\dfrac{\mathcal{M}}{4}-i\Omega & -\dfrac{\mathcal{H}}{4} &
				\dfrac{\big(\mathcal{M}+8i\Omega\big)}{4} & -i\Delta -\dfrac{3\mathcal{H}}{4}-(\gamma + \dfrac{ \gamma _{sg}}{2}) &
				\dfrac{\mathcal{M}}{4} & \dfrac{\mathcal{N}}{2} \\[5pt]
				
				\dfrac{i}{2}\Im(\mathcal{H}) & \dfrac{\mathcal{M}}{4} & 0 &
				\dfrac{3\mathcal{M}^*}{4}-i\Omega & \dfrac{\big(2\mathcal{H}+\mathcal{H}^*\big)}{4} &
				\dfrac{\mathcal{M}^*}{4} &\dfrac{ -\gamma _{sg}-\Re(\mathcal{H})}{2} & \dfrac{3\mathcal{M}}{4}+i\Omega \\[5pt]
				
				\dfrac{\big(\mathcal{M}^*-2i\Omega\big)}{2} & -\dfrac{\mathcal{H}^*}{4} &
				\dfrac{\mathcal{M}^*}{4} & \dfrac{\mathcal{N}^*}{2} &
				\dfrac{\big(\mathcal{M}^*-8i\Omega\big)}{4} & \dfrac{\mathcal{N}^*}{2} &
				(-\dfrac{\mathcal{M}^*}{4}+i\Omega) &( i\Delta -\dfrac{3\mathcal{H}^*}{4}-(\gamma +\dfrac{ \gamma _{sg}}{2}))
			\end{pmatrix}
		\end{aligned}
		$}
\end{equation}
\endgroup 
\end{widetext}
In summary, the derived master equation (Eq.\eqref{x51}) incorporates the effects of the stochastic field through the operators $Z_\pm$, introducing incoherent damping, pumping, and quantum interference terms that modify the atomic dynamics.\\
These contributions, arising from the Born-Markov approximation applied to the weak fluctuating field, enable a detailed analysis of the system's evolution under realistic noisy conditions, as explored in the following sections.\\
Unless otherwise specified, all numerical calculations and plots in this work use the fixed values $\kappa = 60$, $\gamma = 1$, and $\gamma_{sg} = 10^{-3}$, while other parameters (e.g., $\Omega$, $\Delta$, $D$, $\eta$) are varied as indicated in the figures and text. All parameters are given in units of MHz.
\section{Population levels\label{sec:steady}}
We investigate the steady-state population distribution of a $\Lambda$-type  atom driven by coherent and finite-bandwidth stochastic fields. Previous studies (see \cite{Zhou1998,Li2001,Li2002}) demonstrated that stochastic fields can modify level populations and spectral properties, inducing effects such as spectral line narrowing and probe-field amplification. Recent work further indicates that squeezed reservoirs can suppress quantum noise in $\Lambda$-type lasers, enhancing stability \cite{Getahun2025}. Here we compute the steady-state populations by solving Eq.\eqref{x71} using both the bare and dressed-state approaches, and then examine the effects of amplitude and phase fluctuations on population transfer.
\subsection{Bare state}
To obtain the level-population distribution in the bare-state basis, we solve Eq.\eqref{x71} in the steady-state regime. Hence
\begin{equation}
	\rho_A^{{(st)}} = -Q^{-1}B
		\label{x72}
\end{equation}
where the superscript “$st$” denotes the steady state. The steady-state populations are then evaluated numerically from this relation.\\
	\begin{figure}[H]
		\centering
		\subfloat[]{\includegraphics[width=0.98\linewidth]{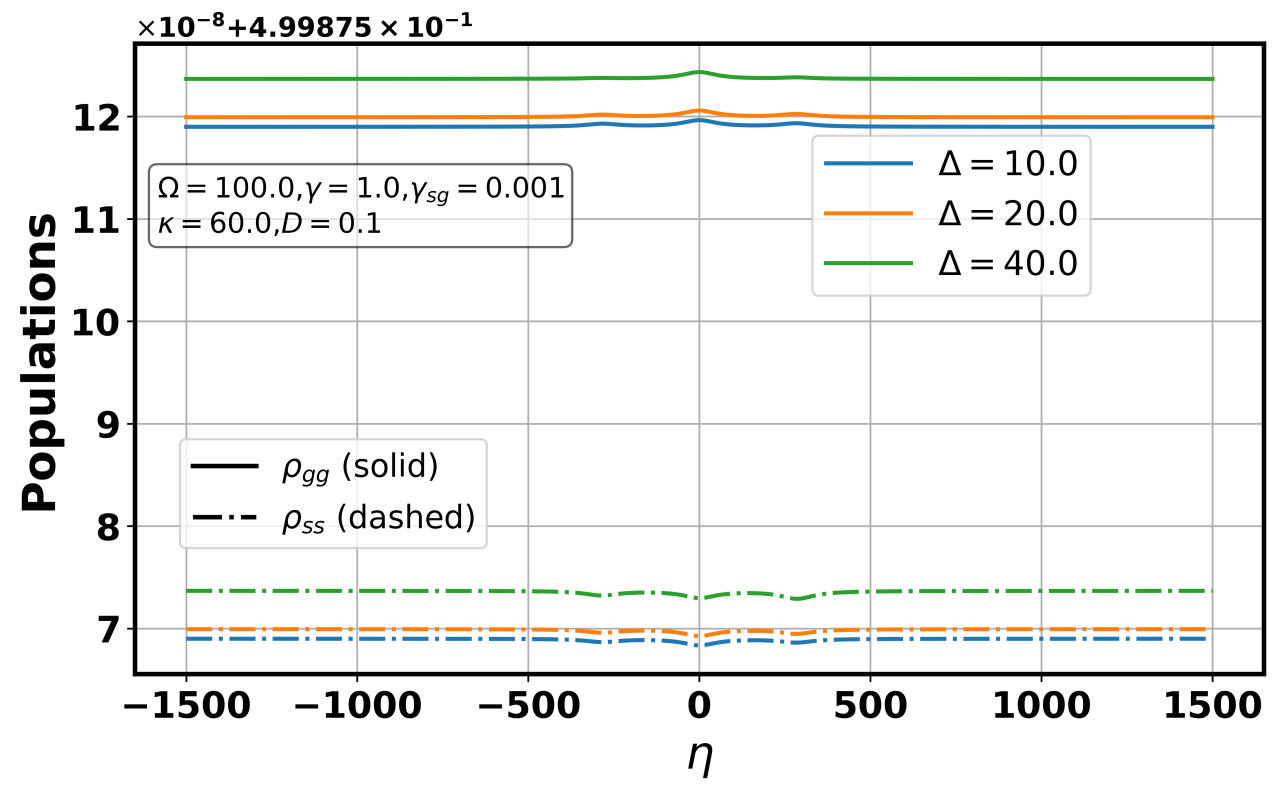}\label{gsd0}}
		\hspace{0.02\linewidth}
		\subfloat[]{\includegraphics[width=0.98\linewidth]{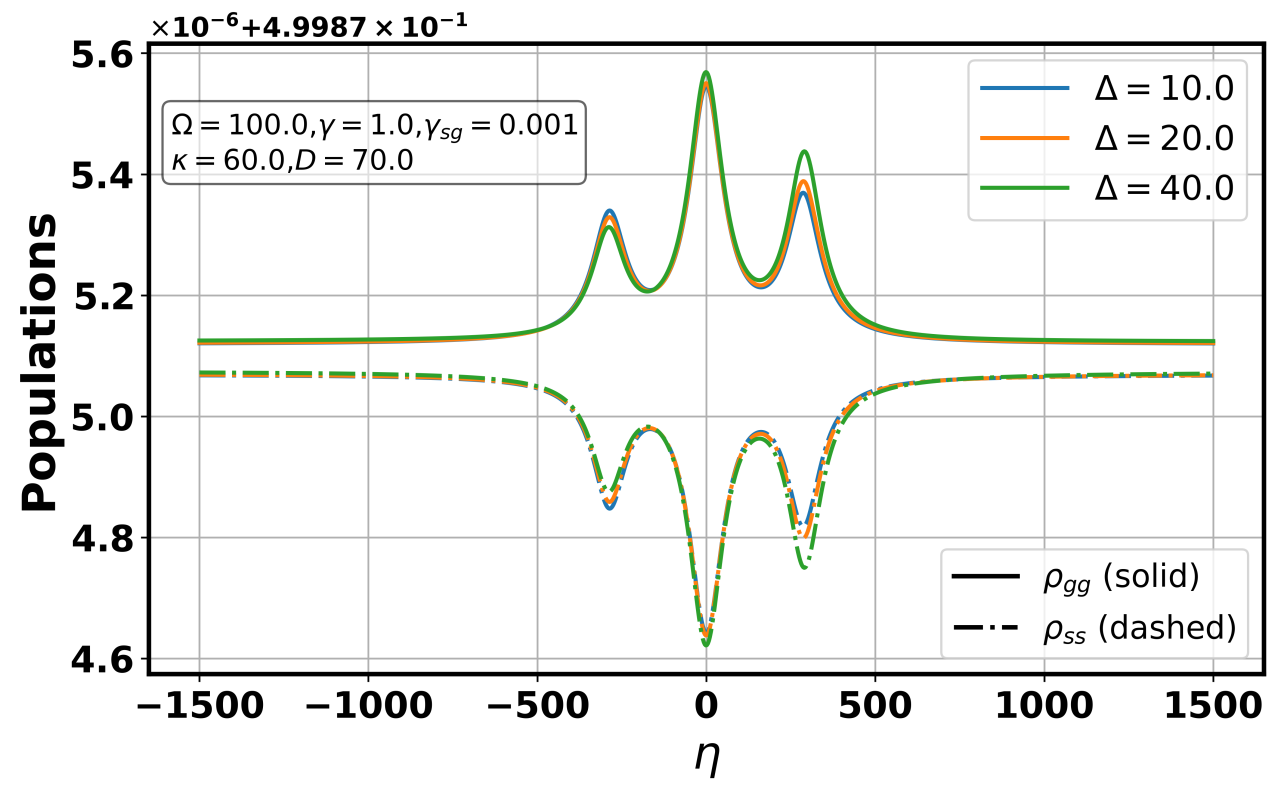}\label{gsd70}}
		
		\caption{Panels (a) and (b) show the steady-state populations of states \(\ket{g}\) and \(\ket{s}\) as functions of the single-photon detuning \(\eta=\omega_s-\omega_L\) for various stochastic-field strengths \(D\). The Rabi frequency is \(\Omega=100\). Solid lines denote \(\rho_{gg}^{(st)}\), and dashed lines denote \(\rho_{ss}^{(st)}\).}
	\label{gsd:main}
	\end{figure}
Fig.\ref{gsd:main} illustrates the steady-state populations of the two ground states, $\rho_{gg}^{(st)}$ and $\rho_{ss}^{(st)}$, as functions of the frequency difference $\eta = \omega_s - \omega_L$ for various noise strengths $D$. In the near-noiseless limit ($D \approx 0$, Fig.\ref{gsd0}), the peaks are narrow and symmetric. As $D$ increases to 70 (Fig.\ref{gsd70}), the central peak broadens into a Lorentzian shape, and its height increases, indicating a transition toward semiclassical behavior. The noisy field influences  Eq.\eqref{x71} through:
$$R(n) = \Re\{f(n)\} = \frac{D \kappa^2}{\kappa^2 + (\eta + n R)^2}.$$
Since $R(n) \propto D$, increasing $D$ amplifies noise-induced transition rates, enhancing the effective ground-state decoherence rate while also strengthening both central and sideband peaks.
As shown in Sec.\ref{pop}, the dressed-state picture of the driven $\Lambda$-type system has three eigenstates: two bright states split by the generalized Rabi frequency and one dark state, yielding a population distribution with up to three peaks.

The single-photon detuning $\Delta$ and Rabi frequency $\Omega$ govern the dressed-state structure, determining the positions and symmetry of the sideband peaks. A nonzero $\Delta$ shifts the dressed-state energies, breaking sideband symmetry and redistributing peak heights, while increasing $\Omega$ enhances the prominence of sideband peaks. Stochastic noise further modifies transition rates and energy shifts, causing sideband peaks to broaden, shift in frequency, or increase in amplitude. Analytical and numerical results confirm these effects.

For small $D$ (Fig.\ref{gsd0}), narrow peaks appear at $\eta\approx\pm R$. For large $D$ (Fig.\ref{gsd70}) the influence of $\Delta$ diminishes, and the population curves converge. This behavior is important for schemes that rely on adiabatic elimination of the intermediate level, where the detuning must be chosen with due consideration of the noise strength $D$.\\
\begin{figure}[H]
	\centering
	\subfloat[]{\includegraphics[width=0.98\linewidth]{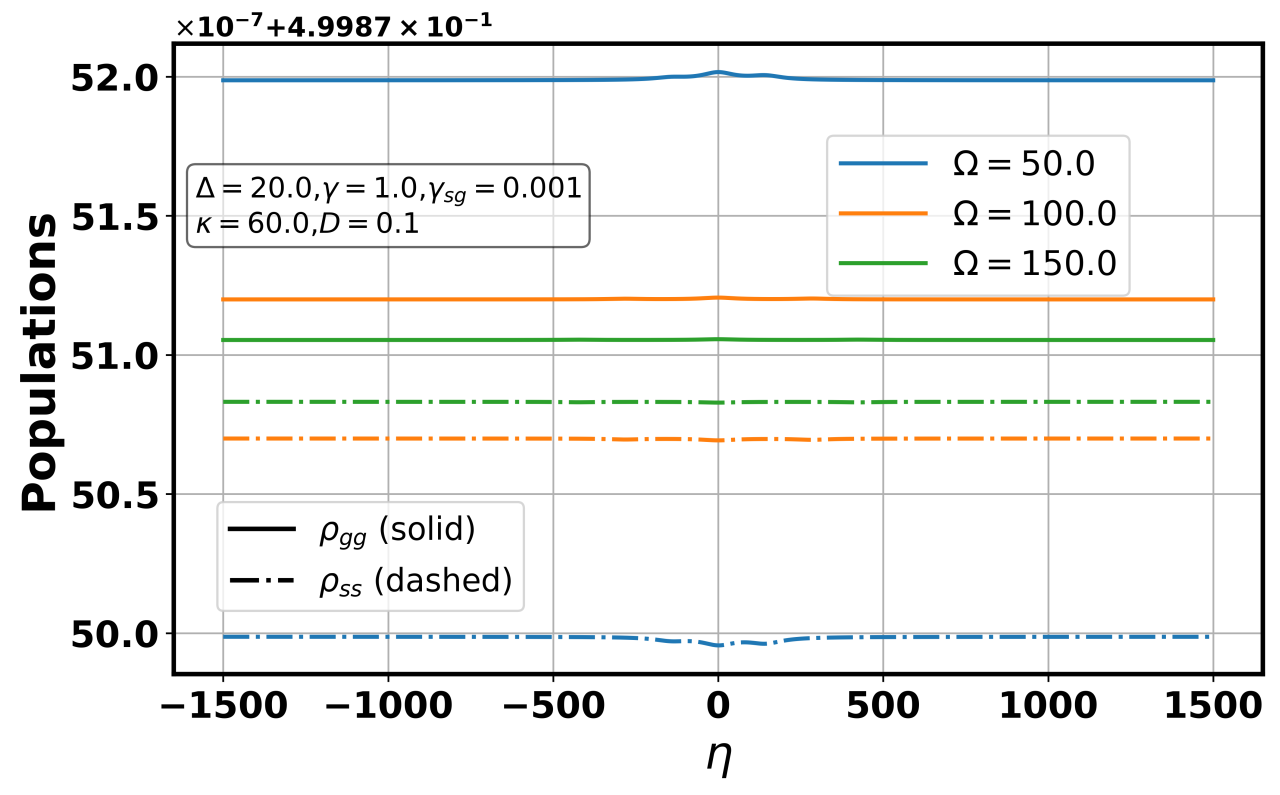}\label{gso0}}\\
	\subfloat[]{\includegraphics[width=0.98\linewidth]{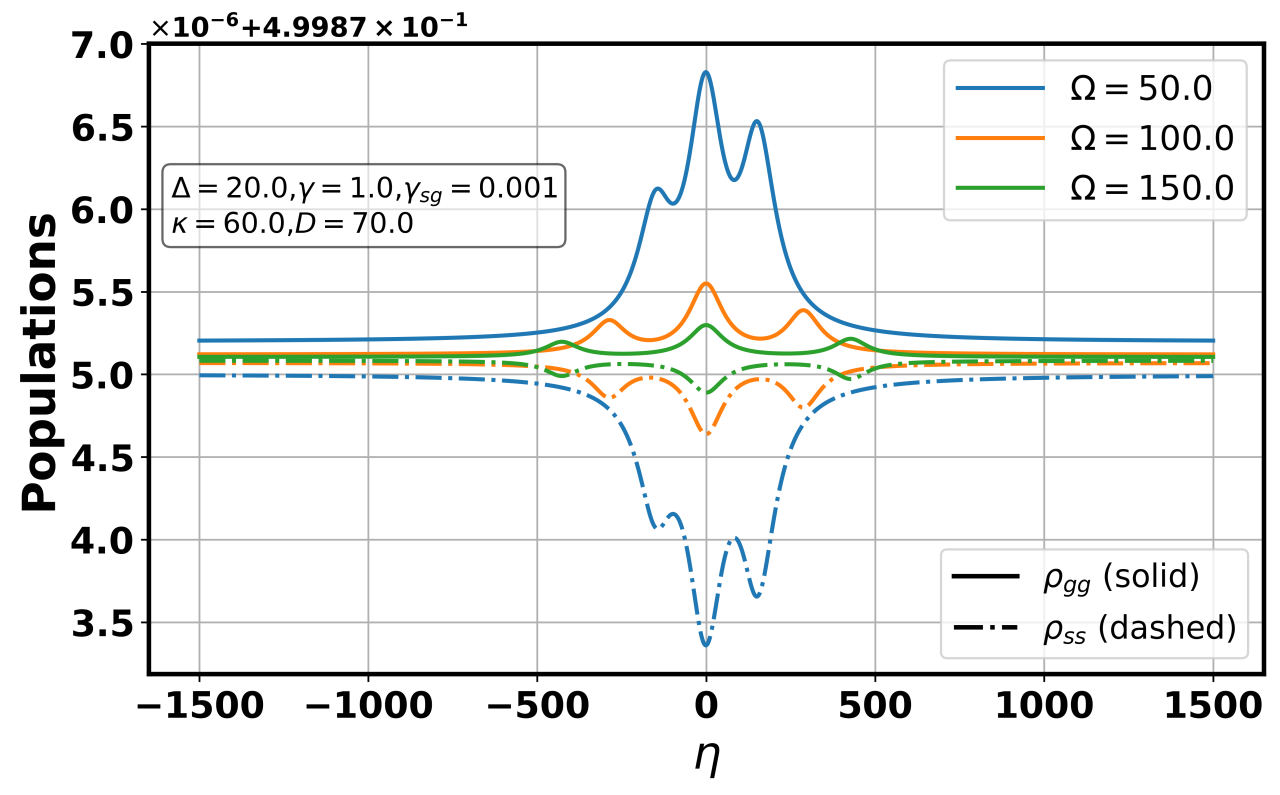}\label{gso70}}
	
	\caption{Panels (a) and (b) show the steady-state populations of the states  \(\ket{g}\) and \(\ket{s}\) for several values of the Rabi frequency \(\Omega\) and stochastic-field strength \(D\), plotted versus the frequency difference \(\eta=\omega_s -\omega_L\) between the coherent and stochastic fields. As in the previous figure, \(\rho_{gg}^{(st)}\) is shown with solid lines and \(\rho_{ss}^{(st)}\) with dashed lines. The single-photon detuning is \(\Delta=20\). Other parameters are given in the text.}
	\label{gso:main}
\end{figure}
Fig.\ref{gso:main} complements the previous figure by again showing the dependence of the steady-state populations on the fluctuating-field frequency \(\eta\). Here, the effects of field fluctuations are examined for increasing fluctuation strengths \(D\) and different Rabi frequencies \(\Omega\). At fixed \(D\), increasing \(\Omega\) suppresses the population of the ground state \(\ket{g}\) while enhancing that of \(\ket{s}\). This manifests in the figure as reduced heights of the \(\rho_{gg}^{(st)}\) peaks and shallower \(\rho_{ss}^{(st)}\) troughs when comparing panels with the same \(D\). However, at fixed \(\Omega\) (e.g., \(\Omega = 50\)), raising \(D\) from near zero to 70 reverses this trend: noise enhances \(\rho_{gg}^{(st)}\) and suppresses \(\rho_{ss}^{(st)}\). This reversal is evident when tracing the evolution of the curves across panels of increasing \(D\).

\begin{figure}[H]
	\centering
	\subfloat[]{\includegraphics[width=0.98\linewidth]{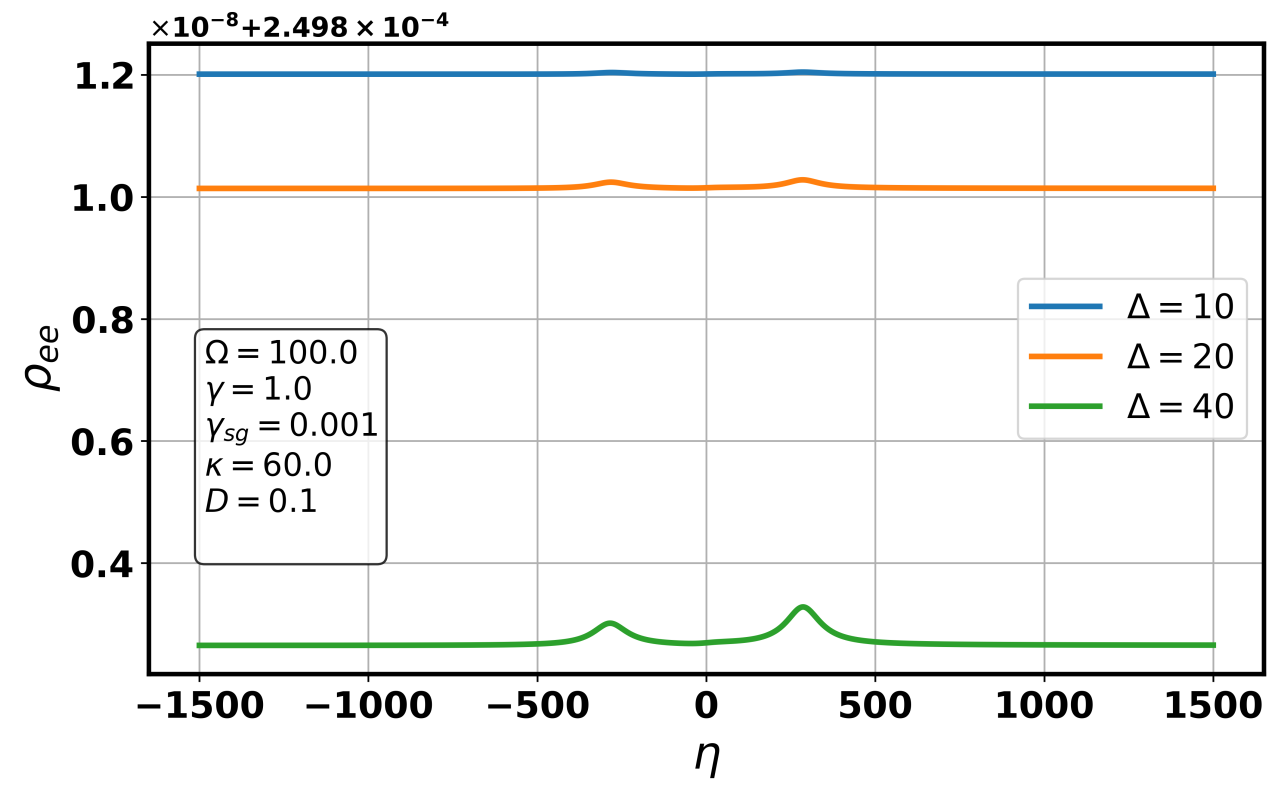}\label{eed0}}\\
	\subfloat[]{\includegraphics[width=0.98\linewidth]{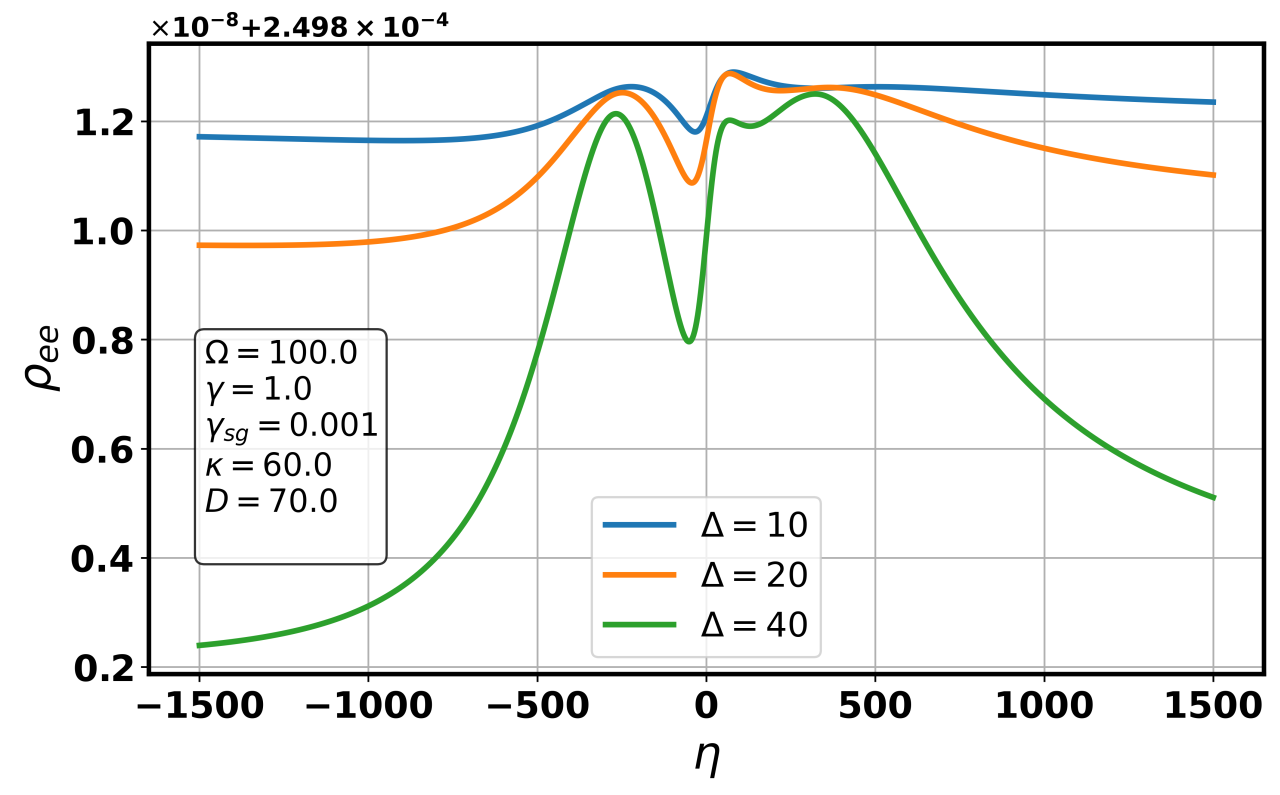}\label{eed70}}
	\caption{Panels (a) and (b) show the steady-state population of the excited state \(\ket{e}\) as a function of the frequency difference \(\eta=\omega_s-\omega_L\) between the coherent and stochastic fields, for several stochastic-field strengths \(D\) and detunings. The Rabi frequency is $\Omega = 100$. }
	\label{eed:main}
\end{figure}

\begin{figure}[H]
	\centering
	\subfloat[]{\includegraphics[width=0.98\linewidth]{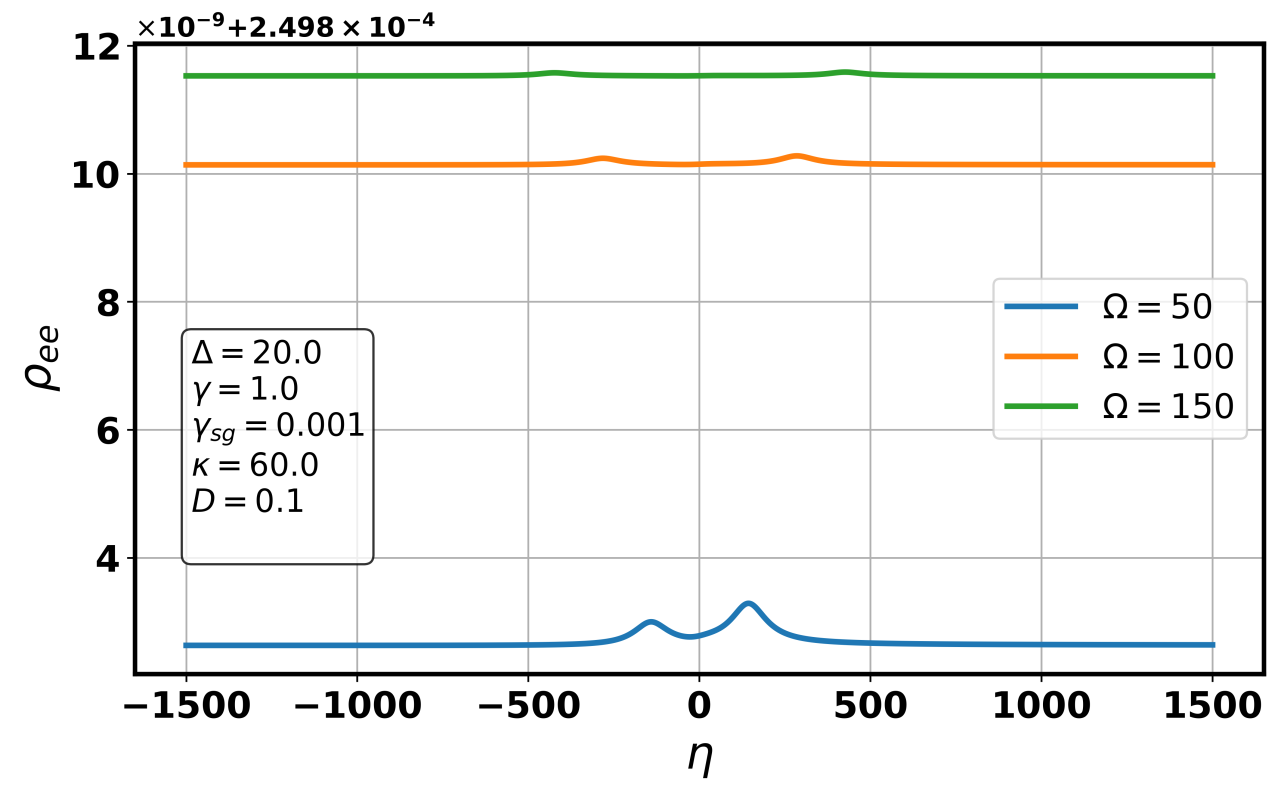}\label{eeo0}}\\
	\subfloat[]{\includegraphics[width=0.98\linewidth]{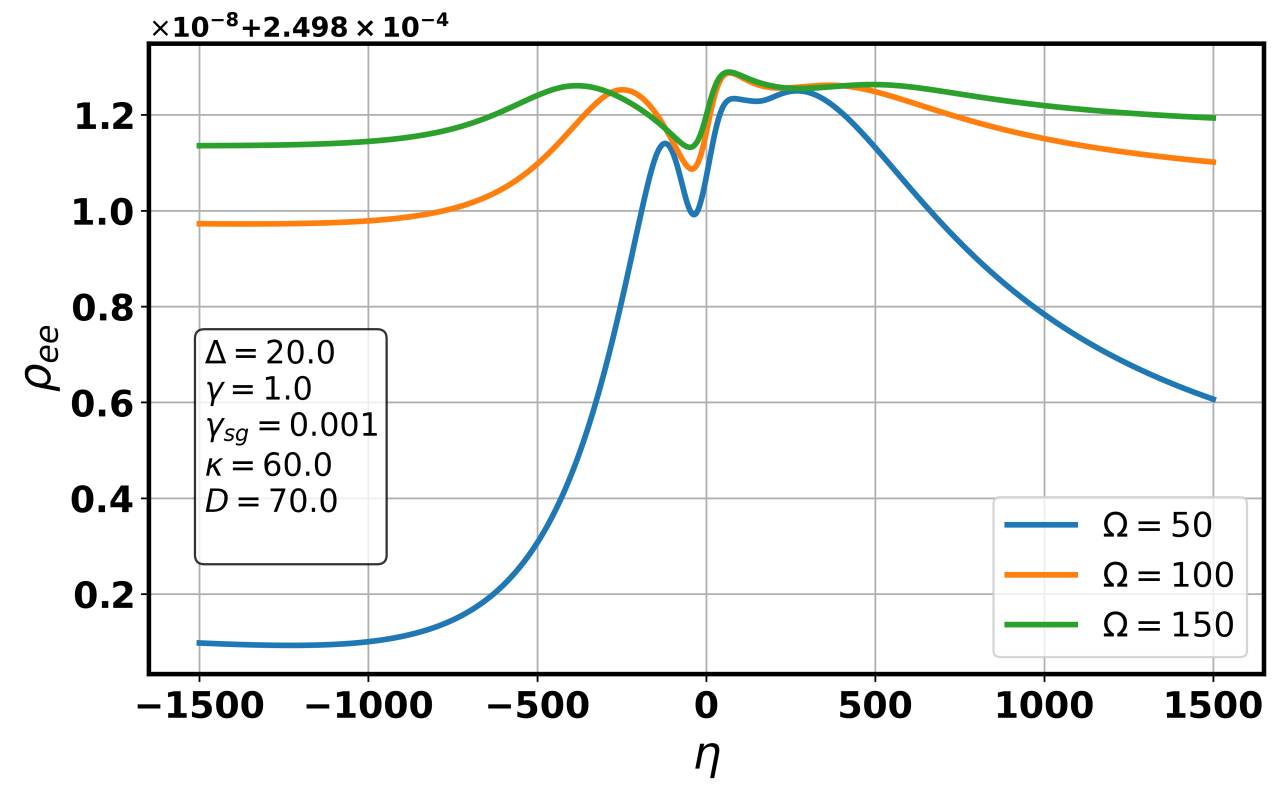}\label{eeo70}}
	\caption{Panels (a) and (b) show the steady-state population of the excited state \(\ket{e}\) as a function of the frequency difference \(\eta=\omega_s-\omega_L\) between the coherent and stochastic fields, for several stochastic-field strengths \(D\) and Rabi frequencies. The single-photon detuning is $\Delta = 20$.}
	\label{eeo:main}
\end{figure}

\begin{figure}[H]
	\centering
     \subfloat[]{\includegraphics[width=0.98\linewidth]{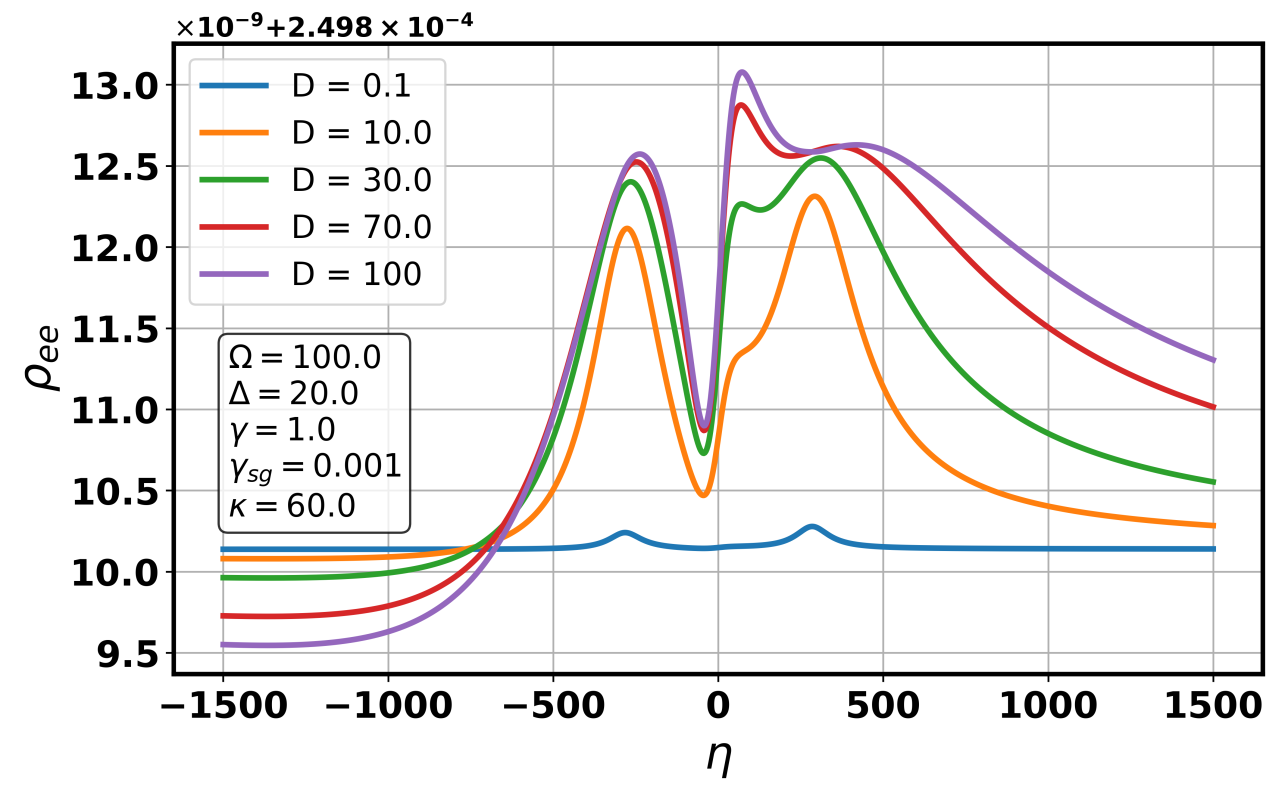}\label{dee}}\\
	\subfloat[]{\includegraphics[width=0.98\linewidth]{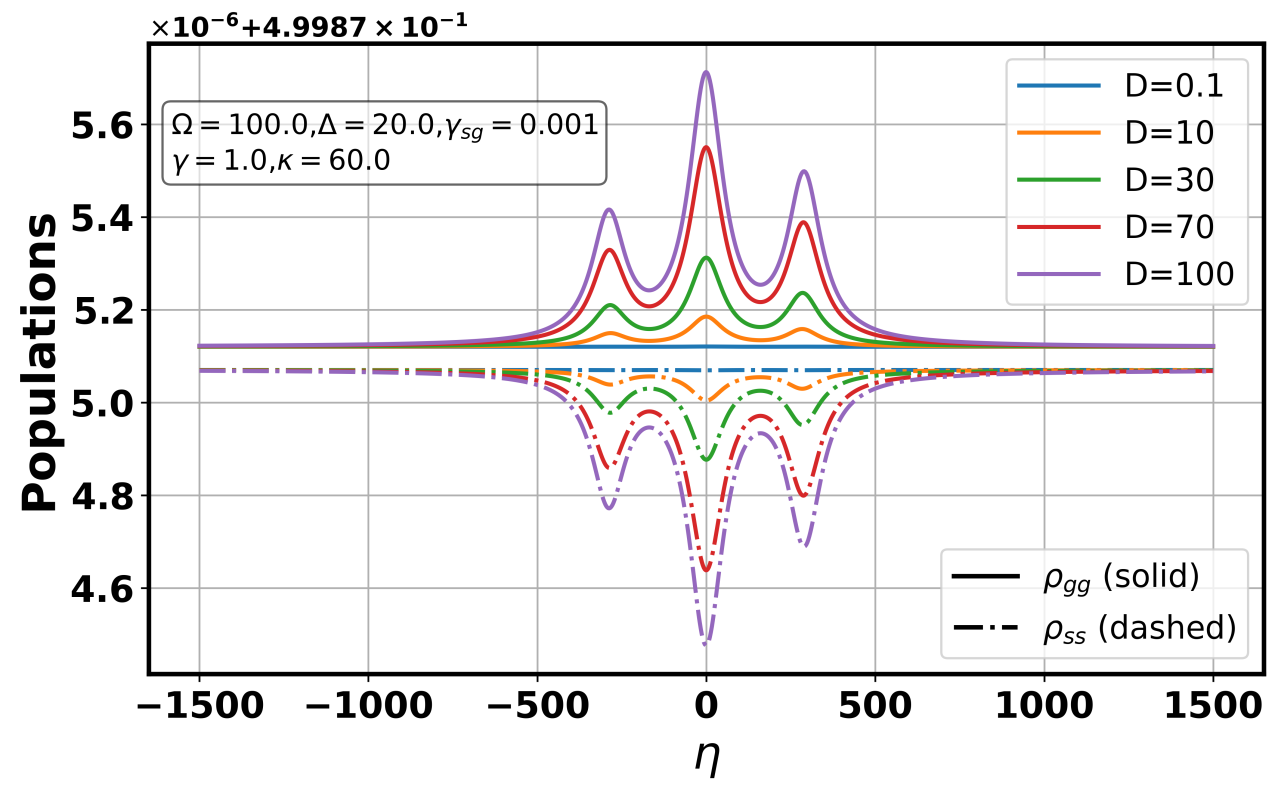}\label{dgs}}
	\caption{Panels (a) and (b) show, respectively, the steady-state populations \(\rho_{gg}^{ (st)}\) and \(\rho_{ss}^{ (st)}\) of the ground states \(\ket{g}\) and \(\ket{s}\), and the steady-state population \(\rho_{ee}^{ (st)}\) of the excited state \(\ket{e}\), plotted as functions of \(\eta=\omega_s-\omega_L\) for several noise strengths \(D\). The Rabi frequency is $\Omega = 100$.}
	\label{D:main}
\end{figure}

In Figs.\ref{eed:main} and \ref{eeo:main} we analyze the intermediate-level population for various fluctuation strengths, detunings, and Rabi frequencies. Similar to the ground-state plots, these figures show that the \(\ket{e}\) population also depends on \(\eta\). However, the steady-state population in \(\ket{e}\) remains small compared with those of \(\ket{g}\) and \(\ket{s}\).
In Fig.\ref{eed:main}, for a fixed fluctuation strength \(D\), increasing the single-photon detuning from \(\Delta = 10\) to \(\Delta = 40\) reduces the population of the intermediate level \(\ket{e}\), with the decrease most pronounced near \(\eta \simeq 0\) and \(\eta \simeq \pm R\). Correspondingly, Fig.\ref{gsd:main} shows that the populations of the ground and metastable states, \(\ket{g}\) and \(\ket{s}\), increase. Thus, a larger \(\Delta\) drives the system toward adiabatic elimination of \(\ket{e}\); however, increasing the fluctuation strength \(D\) counteracts this effect and delays the elimination. This trade-off must be considered when reducing the three-level system to an effective two-level model by eliminating \(\ket{e}\). A comparison with Fig.\ref{eed70} further shows that the peaks and troughs in the \(\ket{e}\) population merge and flatten as \(D\) increases.
In Fig.\ref{eeo:main} the behavior is different. For various Rabi frequencies, resonances appear near \(\eta\simeq\pm R\), and as the fluctuation strength increases the population distributions tend to shift toward the positive-resonance side. This is most evident in panel \ref{eeo70}, where the \(D=70\) profiles converge across different Rabi frequencies.\\
From the figures, we observe that in the nearly noise-free limit (\(D \simeq 0\)), quantum interference is preserved, leading to dark-state formation near the central peak. This arises because the steady-state excited-state population \(\rho_{ee}^{(ss)}\)—responsible for the central peak from the \(\ket{e}\) level—is negligible compared to those of \(\ket{g}\) and \(\ket{s}\) (see Fig.\ref{dee}).
Consequently, the population remains confined to the ground-state manifold, manifesting CPT. Fig.\ref{dee} also reveals that, without noise, the \(\ket{e}\) population is symmetric about \(\eta = 0\); increasing fluctuation strength disrupts this symmetry. In contrast, the ground-state populations \(\rho_{gg}^{(ss)}\) and \(\rho_{ss}^{(ss)}\) retain symmetry and show enhancements at \(\eta \simeq 0, \pm R\) (see Fig.\ref{dgs}).
As $D$ increases, ground-state coherences are suppressed and the dark-state mechanism weakens. Consequently, the CPT feature in $\rho_{gg}^{(st)}$ broadens and its peak height increases, while the corresponding feature in $\rho_{ss}^{(st)}$ becomes shallower.\\
For the explored range of noise strengths \(D\), stochastic fluctuations are sufficient to partially destroy the dark-state coherence. In this situation, a very small fraction of the population is transiently promoted to \(\ket{e}\) and then returns to the ground states, which alters the shape of \(\rho_{gg}^{ (st)}\) and reduces its central peak. At the same time, the sideband peaks associated with the dressed-state splitting induced by the coherent fields become more pronounced. In other words, increasing noise weakens ground-state coherence and quantum interference while enhancing the dynamical splitting due to strong coherent coupling.
The same reasoning applies to the sideband peaks: for large \(\abs{\eta}\) both \(\rho_{gg}^{ (st)}\)  and \(\rho_{ss}^{ (st)}\)  approach constant values, whose asymptotes are determined by the balance of transition rates.
More generally, quantum noise—characterized by the stochastic-field parameters \(D\), $\eta$ and \(\kappa\)—modifies the steady-state profiles near \(\eta\simeq0,\pm R\): it locally enhances \(\rho_{gg}\) and suppresses \(\rho_{ss}\). Increasing \(D\) raises the effective noise-induced rates, broadens spectral features, and eventually amplifies the central peak until semiclassical-like behavior dominates. Hence, stochastic noise substantially alters interference phenomena such as CPT and constrains the parameter window for adiabatic elimination of \(\ket{e}\).
\subsection{Dressed state}\label{pop}
In this section, using the calculations and results from the previous section, we examine the effect of the stochastic field on the steady-state atomic populations in the semiclassical dressed-state basis.
To simplify the master equation, we employ the secular approximation, which facilitates casting the dynamics in Lindblad form. This approximation is valid when the intrinsic evolution timescale of the system $\tau_A$ is much larger than the reservoir relaxation timescale $\tau_S$ (i.e., $\tau_A \gg \tau_S$), which implies $\Omega \gg \gamma$.\\
The dressed eigenstates of \(H_{\Lambda}\) with eigenvalues \(\lambda_0=0\) and \(\lambda_\pm=\tfrac{1}{2}(\Delta\pm R)\), respectively, are
\begin{equation}
	\begin{aligned}
		&\ket{0}=\dfrac{1}{\sqrt{2}}(\ket{g}-\ket{s})\\[12pt]
		& \ket{\pm}= \dfrac{1}{N_\pm}\Big[\dfrac{\Omega}{\lambda_{\pm}}\ket{g} +\ket{e}+\dfrac{\Omega}{\lambda_{\pm}}\ket{s}\Big]
	\end{aligned}
	\label{x59}
\end{equation}
with \(N_\pm=\sqrt{1+2\Omega^2/\lambda_\pm^2}\). Using Eq.\eqref{x59}, we transform the master equation [see Eq.\eqref{x51}] to the dressed-state basis and, after applying the secular approximation, solve for the steady state under the trace constraint \(\rho_{--}+\rho_{00}+\rho_{++}=1\), yielding the steady populations given in Eq.\eqref{x141}.
\begin{equation}
	\begin{aligned} 
		&\rho _{00}=\frac{4R(\Delta +R) \left(\Gamma _1 \Gamma _4-\Gamma _2 \Gamma _3\right) }{\mathcal{T}} \\[4pt]
		&\rho _{++}=\frac{\gamma _{sg}\left(4 \Gamma _2 \Omega ^2+\Gamma _4 R(\Delta +R)\right)}{\mathcal{T}}\\[4pt]
		& \rho _{--}=-\frac{\gamma _{sg}\left(4 \Gamma _1 \Omega ^2+\Gamma _3 R (\Delta +R)\right)}{\mathcal{T}}
	\end{aligned}
	\label{x141}
\end{equation}
Here \(\rho_{00}\), \(\rho_{++}\) and \(\rho_{--}\) denote the diagonal elements of the density operator in the dressed-state basis. The \(\Gamma\) coefficients (see Eq.\eqref{x140}) represent effective decay rates between dressed levels. 

The coefficients $\Gamma_1$ and $\Gamma_2$ in Eq.\eqref{x140} represent contributions solely from intrinsic decay processes, specifically spontaneous decay from the excited state to the ground states ($\gamma$) and ground-state decoherence ($\gamma_{sg}$). By contrast, $\Gamma_3$ and $\Gamma_4$ contain both an intrinsic decay component and a noise-dependent contribution $\mathcal{C}$ (see Eq.\eqref{x140}) arising from the stochastic-field interaction. The quantity $\mathcal{C}$ is built from the real parts $\operatorname{Re}(\mathcal{M})$, $\operatorname{Re}(\mathcal{N})$, and $\operatorname{Re}(\mathcal{H})$ and therefore depends on $\eta = \omega_s - \omega_L$, the stochastic-field strength $D$, and the bandwidth $\kappa$. \\
This noise-dependent term is responsible for the resonant structures and peak broadening observed in the steady-state populations. The common denominator appearing in Eq.\eqref{x141} is denoted by $\mathcal{T}$.
{\small \begin{equation}
		\begin{aligned}
			&\Gamma_{1} = \frac{2\gamma _{sg} \Omega ^2+\gamma (R+\Delta)^2}{2 R (R+\Delta)}\\[8pt]
			&\Gamma_{2} =\Gamma_{1}  + \frac{\Delta(\gamma _{sg}-4\gamma)}{4R}\\[8pt]
			&\Gamma_{3} = \dfrac{\big(2\Omega^2-R(R+\Delta)\big)\big(\gamma(R+\Delta)^2 + 2\gamma_{sg}\Omega^2\big)}{R^2(R+\Delta)^2}  - \dfrac{\mathcal{C}}{2R^2} \\[8pt]
			&\Gamma_{4} =\dfrac{2\gamma _{sg} \Omega ^2 +\gamma (R-\Delta )^2}{4R^2} + \dfrac{\mathcal{C}}{2R^2}\\[8pt]
			&\mathcal{C}=  Re(\mathcal{H}) \left(\Delta ^2+R^2\right)+8 \Omega  \big(\Delta  Re(\mathcal{M})-\Omega  Re(\mathcal{N})\big)\\[8pt]
			& \mathcal{T}= R(\Delta +R)\Big(4 \left(\Gamma _1 \Gamma _4-\Gamma _2 \Gamma _3\right)+\gamma _{sg}\left(\Gamma _4-\Gamma _3\right)\Big)\\
		    &\qquad+4 \Omega ^2\gamma _{sg}\left(\Gamma _2-\Gamma _1\right)
		\end{aligned}
		\label{x140}
\end{equation}}
In the limiting case of $\kappa \to \infty$ and $D \to 0$ (the ideal noise-free regime), Eq.\eqref{x141} reduce to the following form
\begin{equation}
	\begin{aligned}
		&\rho_{++}=\frac{\gamma _{sg} \left[\Omega ^2 (4 \gamma +\gamma _{sg})-\gamma  \Delta(R  - \Delta )\right]}{4 \gamma \gamma _{sg} \Delta ^2+\Omega ^2 (4 \gamma +\gamma _{sg}) (4 \gamma +3\gamma _{sg})}\\[15pt]
		&\rho_{00}= \frac{2 \gamma \gamma _{sg} \Delta ^2+\Omega ^2 (4 \gamma +\gamma _{sg})^2}{4 \gamma \gamma _{sg} \Delta ^2+\Omega ^2 (4 \gamma +\gamma _{sg}) (4 \gamma +3\gamma _{sg})}\\[15pt]
		&\rho_{--} = \frac{\gamma _{sg} \left[ \Omega ^2 (4 \gamma +\gamma _{sg}) + \gamma  \Delta  \left(R+\Delta \right) \right]}{4 \gamma \gamma _{sg} \Delta ^2+\Omega ^2 (4 \gamma +\gamma _{sg}) (4 \gamma +3\gamma _{sg})} 
	\end{aligned}
	\label{x142}
\end{equation}
We observe all contributions associated with the noise of the fluctuating field have indeed vanished from the above population functions, and a smooth, fluctuation-free distribution in both amplitude and phase emerges for the populations of the dressed states of this $\Lambda$-type atom.\\

In this regime, examining the effect of the decoherence rate $\gamma_{sg}$ provides useful physical intuition. From Eq.\eqref{x142}, it is evident that if $\gamma_{sg} = 0$, then $\rho_{00} = 1$ and the populations of the other levels vanish. Hence, a nonzero value of $\gamma_{sg}$ is required to observe various phenomena, such as redistributed level populations and the incoherent fluorescence spectrum.\\
Applying $\Delta \to 0$ to Eq.\eqref{x142} yields
\begin{equation}
	\begin{aligned}
		&\rho_{00}= \frac{4 \gamma + \gamma_{sg}}{4 \gamma +3\gamma_{sg}}\\
		&\rho_{++} =\frac{\gamma_{sg}}{4 \gamma +3 \gamma_{sg}}\\
		&\rho_{--} =\frac{\gamma_{sg}}{4 \gamma +3 \gamma_{sg}}
	\end{aligned}
	\label{x143}
\end{equation}
\begin{figure}[H]
	\centering
	\subfloat[]{\includegraphics[width=0.98\linewidth]{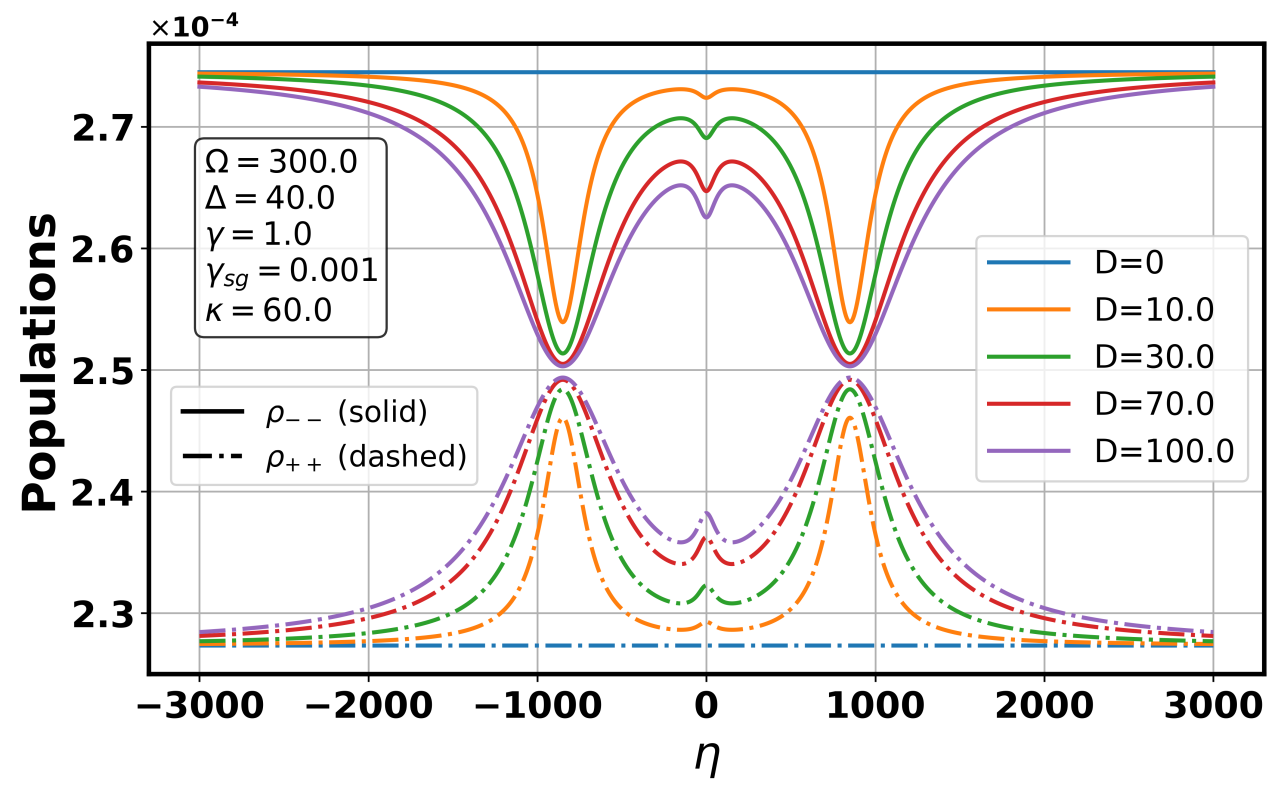}\label{drsdel40}}\\
	\subfloat[]{\includegraphics[width=0.98\linewidth]{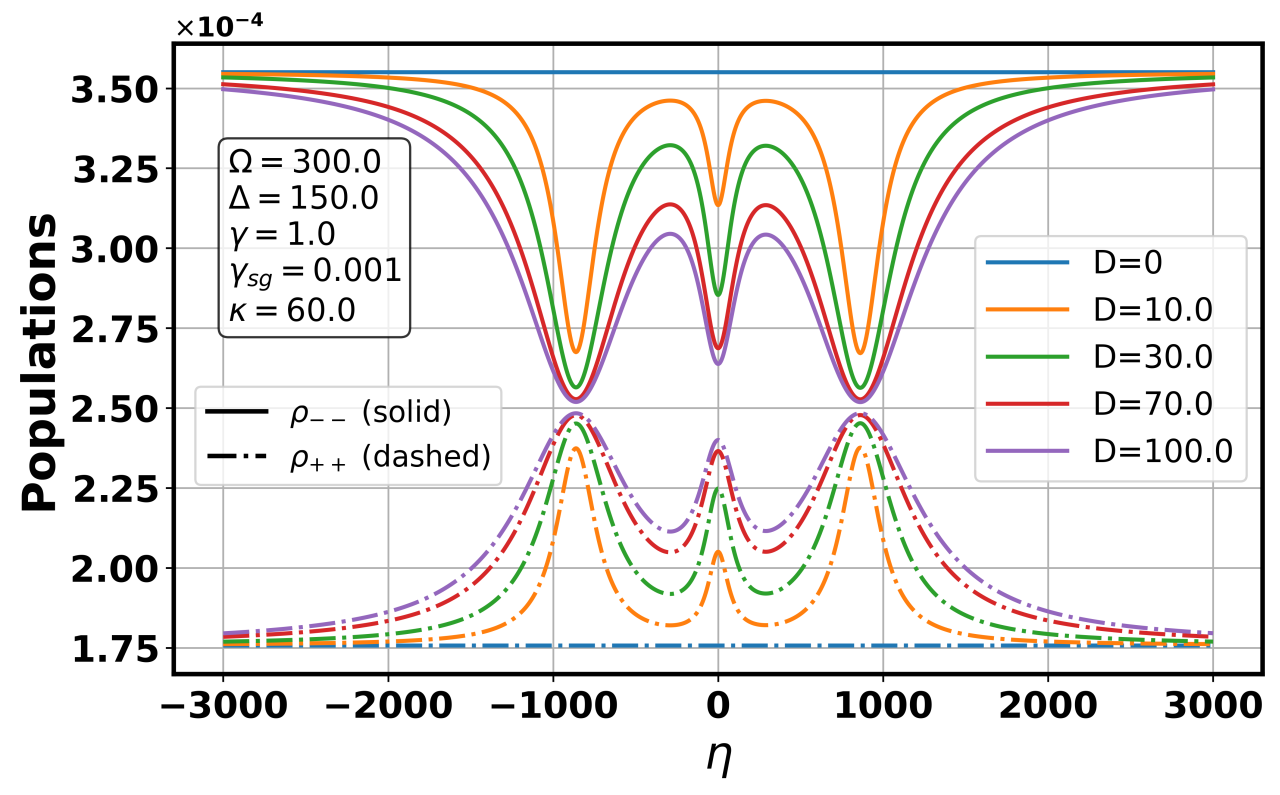}\label{drsdel150}}
\caption{Steady-state populations of the dressed states \( \ket{+} \) and \( \ket{-} \) as functions of \( \eta = \omega_s - \omega_L \) for various stochastic-field strengths \( D \). Panels (a) and (b) correspond to single-photon detunings \( \Delta = 40 \) and \( \Delta = 150 \), respectively. The Rabi frequency is fixed at \( \Omega = 300 \). Solid and dashed lines represent \( \rho_{--}^{(st)} \) and \( \rho_{++}^{(st)} \), respectively.}
	\label{drs1}
\end{figure}
Which implies that population inversion between the dressed states $\ket{\pm}$ never occurs in this regime. However, as soon as the single-photon detuning $\Delta$ deviates from zero, this population equality breaks down; increasing the stochastic-field strength $D$ further reveals resonant features (see Fig.\ref{drs1}).\\
Fig.\ref{drs1} displays the dressed-state steady-state populations \(\rho_{--}^{st}\) and \(\rho_{++}^{st}\), while Fig.\ref{drs2} shows \(\rho_{00}^{st}\).\\
In these plots, the population distribution is uniform for $D=0$, but increasing $D$ leads to resonances near $\eta=0, \pm R$, arising from the stochastic terms $\operatorname{Re}(\mathcal{N})$, $\operatorname{Re}(\mathcal{M})$, and $\operatorname{Re}(\mathcal{H})$ in the equations for $\rho_{--}^{(st)}$, $\rho_{++}^{(st)}$, and $\rho_{00}^{(st)}$. This resonant feature indicates that the atom-field interaction forms a dressed atom whose energy-level structure is strongly field-intensity dependent, with spontaneous emission dominated at the three frequencies $0$ and $\pm R$. 
The dressed atom is then driven by the stochastic field. Consequently, when the central frequency of the stochastic field is tuned to these frequencies, the corresponding atomic transition is enhanced.
\begin{figure}[H]
	\centering
	\subfloat[]{\includegraphics[width=0.99\linewidth]{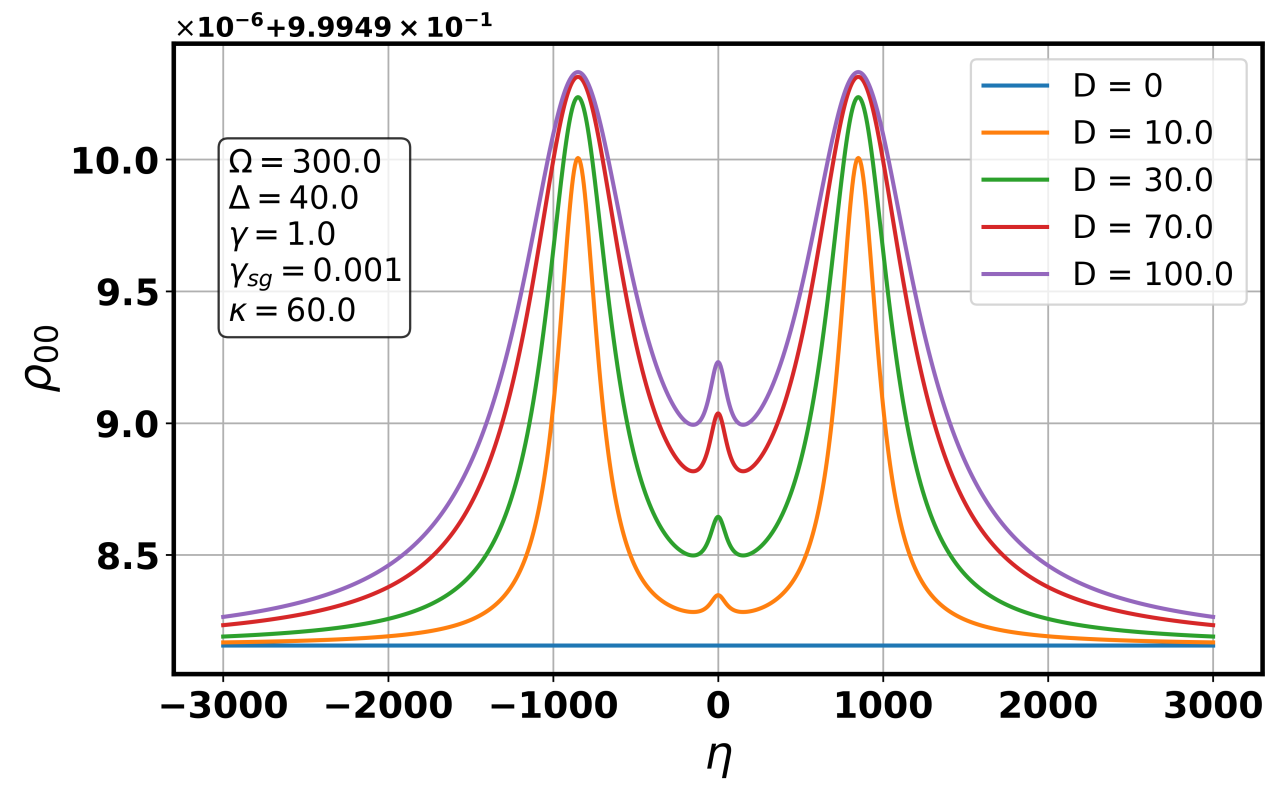}\label{drs0del40}}\\
	\subfloat[]{\includegraphics[width=0.99\linewidth]{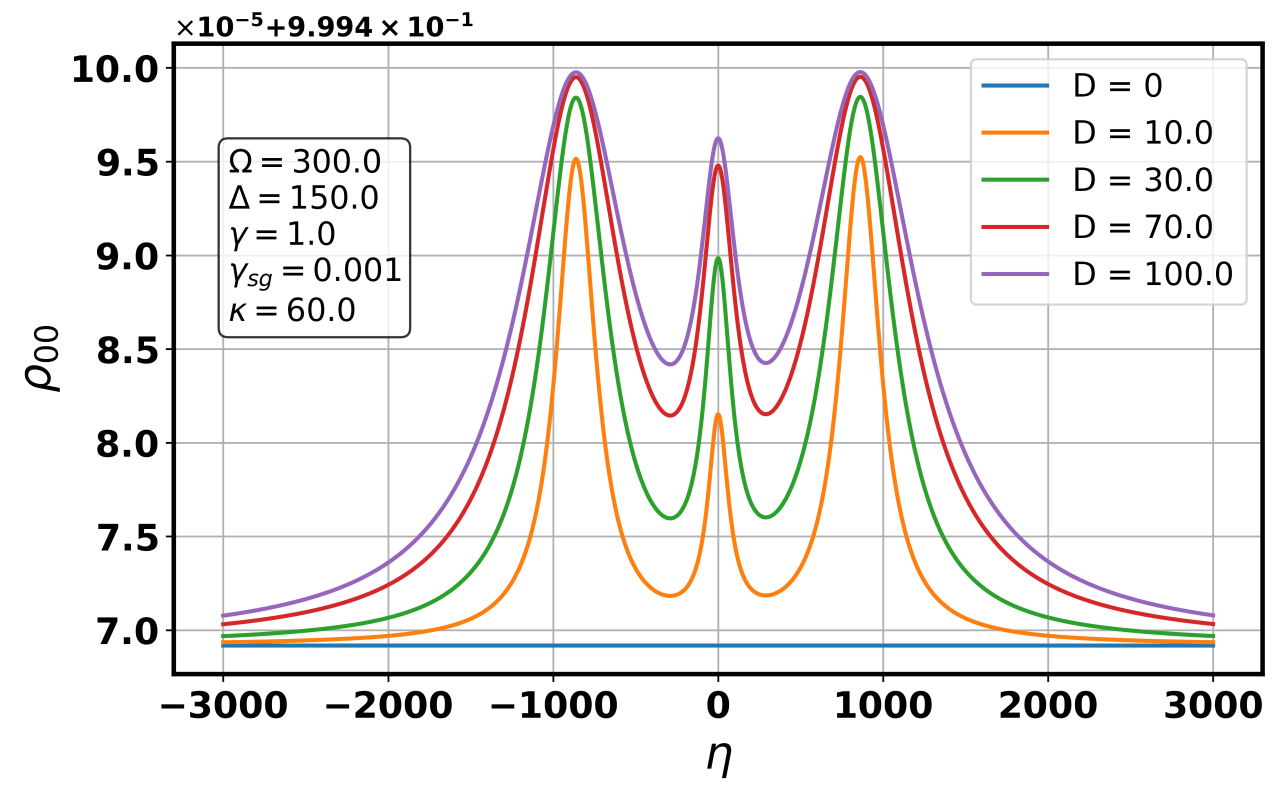}\label{drs0del150}}
    \caption{Steady-state population of the dressed state \(\ket{0}\) as a function of \(\eta = \omega_s - \omega_L\). Panels (a) and (b) correspond to stochastic-field strengths \(D\) at single-photon detunings \(\Delta = 40\) and \(\Delta = 150\), respectively. Parameters are as in Fig.\ref{drs1}.}
	\label{drs2}
\end{figure}
In conclusion, we have shown that a stochastic field — with an appropriately chosen central frequency and finite bandwidth — significantly modifies the dressed-state populations of a \(\Lambda\)-type three-level atom, producing resonant structures in the population distribution. \\
The derived expressions indicate that increasing the fluctuation intensity \(D\) broadens spectral peaks and, depending on the dressed level, either reduces or enhances their amplitudes. \\
In the noise-free limit these fluctuation-induced effects vanish and the populations are governed solely by the coherent dynamics. The changes originate from noise-dependent contributions to the effective decay rates between dressed levels and thus elucidate the role of stochastic fields in the system's steady-state dynamics.\\

Finally, although the secular approximation is well justified in the strong-field regime and greatly simplifies the analysis, it should be applied with care since it may neglect interference terms that are important in certain parameter regimes.
\section{Incoherent Resonance Fluorescence\label{sec:fluorescence}}
In the ideal case, a two-level atom driven by a strong, perfectly coherent laser field emits the well-known three-peaked Mollow spectrum.\\
In practice, however, no laser is perfectly monochromatic; realistic laser fields exhibit phase fluctuations and finite linewidth. The resulting incoherent resonance fluorescence modifies the Mollow structure—in particular, laser phase noise substantially broadens the spectral peaks, with the central (Rayleigh-like) peak being especially sensitive to such fluctuations. Accounting for these effects improves agreement between theory and experiment and shows how source noise and incoherence can alter the quantum features of the system.

In this section we analyze the incoherent resonance-fluorescence spectrum. The spectrum is defined by \cite{zub}
\begin{equation}
	S(\omega) = Re\int_{0}^{\infty}\lim_{t\to\infty} G^{(1)}(\tau) e^{i\omega\tau}d\tau
	\label{r1}
\end{equation}
where \(G^{(1)}(t,\tau)\) is the first-order two-time correlation function \cite{Walls}. For the three-level \(\Lambda\) atom with dipole moments \(\boldsymbol{\mu}_{eg}\) and \(\boldsymbol{\mu}_{es}\) one has
\begin{equation}\label{r3}
	\begin{aligned}
		G^{(1)}(t,\tau)&=|\boldsymbol{\mu}_{es}|^2\langle\sigma_{es}(t)\sigma_{se}(t+\tau)\rangle\\
		&+|\boldsymbol{\mu}_{eg}|^2\langle\sigma_{eg}(t)\sigma_{ge}(t+\tau)\rangle\\
		&+\boldsymbol{\mu}_{es}\!\cdot\!\boldsymbol{\mu}_{eg}^*\,\langle\sigma_{es}(t)\sigma_{ge}(t+\tau)\rangle\\
		&+\boldsymbol{\mu}_{eg}\!\cdot\!\boldsymbol{\mu}_{es}^*\,\langle\sigma_{eg}(t)\sigma_{se}(t+\tau)\rangle .
	\end{aligned}
\end{equation}
We assume that the transition dipoles are orthogonal so that the cross terms vanish, \(\boldsymbol{\mu}_{es} \cdot \boldsymbol{\mu}_{eg}^* = \boldsymbol{\mu}_{eg} \cdot \boldsymbol{\mu}_{es}^* \simeq 0\), and, for simplicity, \(|\boldsymbol{\mu}_{es}|^2 = |\boldsymbol{\mu}_{eg}|^2 \equiv |\boldsymbol{\mu}|^2\).  Decomposing each atomic transition operator into its steady-state expectation and a fluctuation term, $	\sigma_{ij} = \expval{\sigma_{ij}}_{st} + \delta \sigma_{ij}(t)$, yields the incoherent spectrum from the two-time correlations of the fluctuations \cite{meystre}
\begin{equation}
	S_{inc}(\omega) = \abs{ \boldsymbol{\mu}}^2  Re \sum_{k=g,s} \int_{0}^{\infty} G_k^{(1)}(\tau) e^{i\omega\tau}d\tau 
	\label{r8}
\end{equation}
where \( G_k^{(1)}(\tau) = \langle \delta\sigma_{ek}(t)\delta\sigma_{ke}(t+\tau) \rangle_{\text{st}} \). This spectrum must be calculated using the quantum regression theorem (QRT) and Laplace transform techniques.\\
Evaluating \( G_k^{(1)}(\tau) \) reveals that each correlator \( \langle \delta\sigma_{ij} \rangle \) is coupled to others. To compute Eq.\eqref{r8}, we therefore define a new vector. Setting \( t=0 \) and using the relation \( \langle \sigma_{ij} \rangle = \rho_{ji} \) in conjunction with Eq.\eqref{xx70}, we construct a column vector of the required two-time correlations:
\begin{equation}
	y_k(\tau)=
	\begin{pmatrix}
		\expval{\delta \sigma_{ek}(0)\delta \sigma_{gg}(\tau)}_{st}\\
		\expval{\delta \sigma_{ek}(0)\delta \sigma_{eg}(\tau)}_{st}\\
		\expval{\delta \sigma_{ek}(0)\delta \sigma_{sg}(\tau)}_{st}\\
		\expval{\delta \sigma_{ek}(0)\delta \sigma_{ge}(\tau)}_{st}\\
		\expval{\delta \sigma_{ek}(0)\delta \sigma_{ee}(\tau)}_{st}\\
		\expval{\delta \sigma_{ek}(0)\delta \sigma_{se}(\tau)}_{st}\\
		\expval{\delta \sigma_{ek}(0)\delta \sigma_{gs}(\tau)}_{st}\\
		\expval{\delta \sigma_{ek}(0)\delta \sigma_{es}(\tau)}_{st}\\
	\end{pmatrix},\quad k=g,s.
	\label{r17}
\end{equation}
Alternatively, \( G_k^{(1)}(\tau) \) can be calculated by applying the QRT to a vector of single-time expectation values. The steady-state single-time expectation vector is
 \begin{equation}
 	\begin{aligned}
 		\expval{X(t)} =& \big[\expval{\sigma_{gg}(t)},\expval{\sigma_{eg}(t)},\expval{\sigma_{sg}(t)},\expval{\sigma_{ge}(t)},\\
 		&\expval{\sigma_{ee}(t)},\expval{\sigma_{se}(t)},\expval{\sigma_{gs}(t)},\expval{\sigma_{es}(t)}\big]^T
 	\end{aligned}
 	\label{r9}
 \end{equation}
The dynamics of the fluctuation vector \( \langle \delta \mathbf{X} \rangle \) are then governed by the following equation for the time evolution of the fluctuations, which is valid in the context of Eq.\eqref{x71}
\begin{equation}
	\dfrac{d}{d\tau}\expval {\delta X}=Q\expval{\delta X}
	\label{r15}
\end{equation}
According to QRT, the two-time correlation vector in Eq.\eqref{r17} satisfies the same homogeneous equation, Eq.\eqref{r15}.
\begin{equation}
	\dfrac{dy_k(\tau)}{d\tau}=Qy_k(\tau)
	\label{r18}
\end{equation}
Taking the Laplace transform of Eq.\eqref{r18} gives
	\begin{equation}
	Y_{k}(\omega) =-M y_k(0), \ M=(i \omega I + Q)^{-1}, \ k=g,s
	\label{r23}
\end{equation} 
Eq.\eqref{r23} holds when all eigenvalues of \(i\omega I+Q\) have negative real parts, i.e. $\Re\left[\lambda(i\omega I+Q)\right]<0\quad\forall\ \omega$. Hence the Laplace integral converges uniformly in \(\omega\), so
\begin{equation}
	S_{inc}(\omega)=\abs{ \boldsymbol{\mu}}^2  Re\left[ Y^{(4)}_{g}(\omega) + Y^{(6)}_{s}(\omega)\right]
	\label{r21}
\end{equation}
The initial correlation vectors $y_g(0)$ and $y_s(0)$ in Eq.\eqref{r21} can be calculated using Eq.\eqref{r17} for $\tau=0$ and $\sigma_{ij} = \langle\sigma_{ij}\rangle_{st} + \delta \sigma_{ij}(t)$.

Eqs.\eqref{r21} and \eqref{r23} together provide the incoherent fluorescence spectrum once the steady-state means and the matrix \(Q\) are known (numerically or analytically). In particular, the noise-dependent modifications to the steady-state means (via \(\mathcal{M},\mathcal{N},\mathcal{H}\), and the \(\Gamma_i\)) enter the spectrum through \(y_k(0)\) and \(Q\), producing the resonant structures and broadenings discussed in the text and shown in the figures.

\begin{figure}[H]
	\centering
	\subfloat[]{\includegraphics[width=0.95\linewidth]{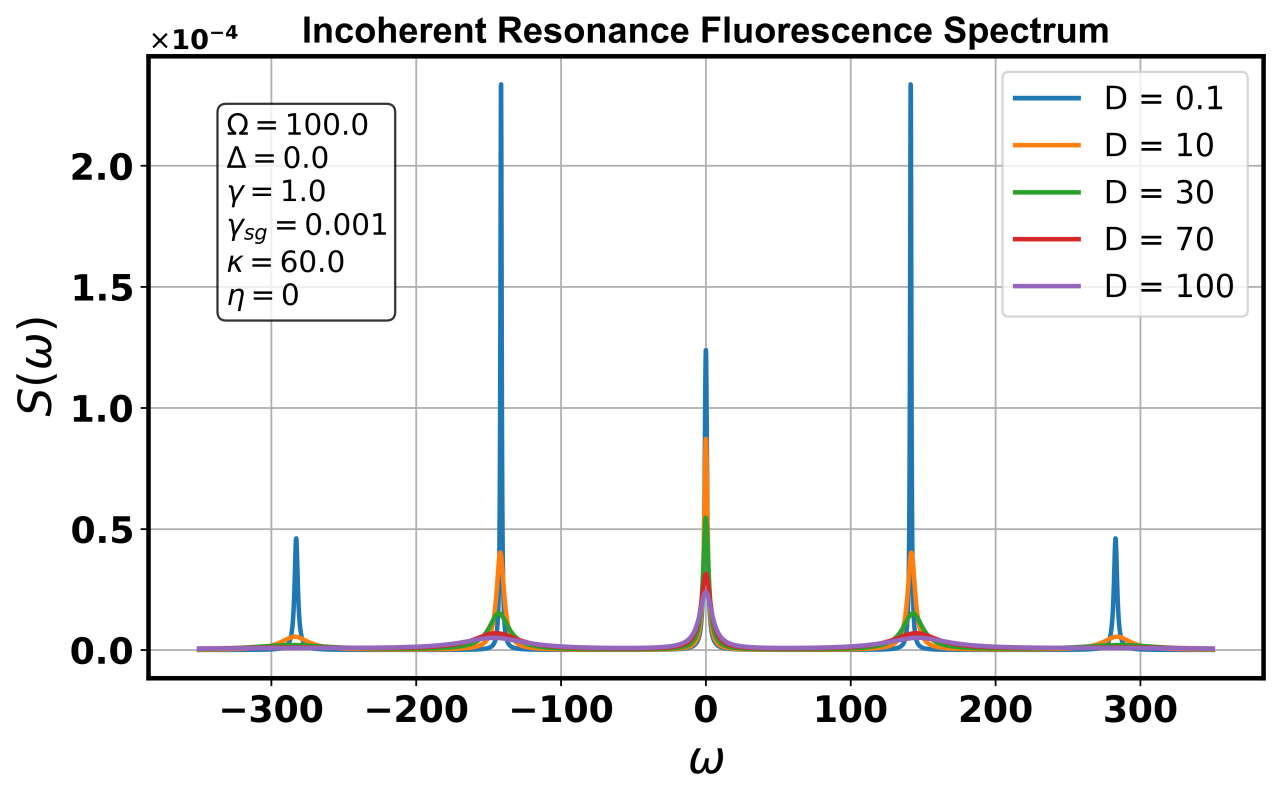}\label{resdel0}}\\
	\subfloat[]{\includegraphics[width=0.95\linewidth]{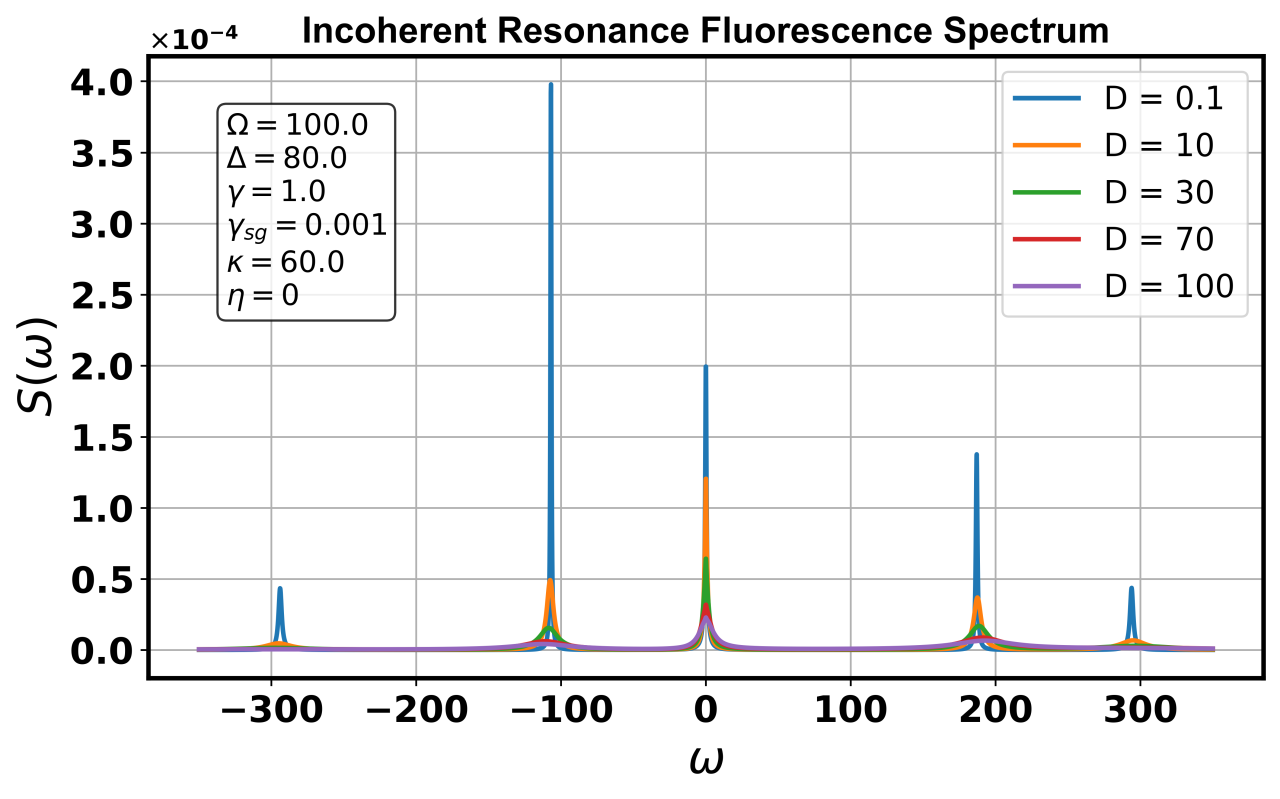}\label{resdel80}}
	\caption{Panels (a) and (b): Incoherent resonance-fluorescence spectrum $S_{\rm inc}(\omega)$ for stochastic-field strengths $D=0.1,\,10,\,30,\,70$, and $100$ (at $\Delta=0, 80$, respectively) and $\eta=0$. The parameters are $\Omega = 100$ and $\mu = 1$. Other parameters are as specified in the text.}
	\label{res-del}
\end{figure}

Fig.\ref{res-del} shows the incoherent resonance-fluorescence spectrum for various single-photon detunings at $\eta = 0$, illustrating the effects of amplitude and phase fluctuations. Notably, at $\Delta = 0$ (see Fig.\ref{resdel0}), the spectrum evolves into five symmetric peaks as the fluctuation strength $D$ increases through the values $0.1, 10, 30, 70,$ and $100$.
However, for $\Delta =  80$ (see Fig.\ref{resdel80}), the spectrum loses symmetry. Increasing \(\Delta\) thus alters peak heights and introduces asymmetry, arising from changes in coherence fractions and quantum interactions. This loss of spectral symmetry is a direct consequence of the detuning-induced imbalance in dressed-state populations (as detailed in Sec.\ref{pop}). Since each spectral peak's intensity depends on the initial population of the corresponding dressed state and the transition rates, this population asymmetry manifests as an asymmetric fluorescence spectrum.\\
These figures highlight two key features of the incoherent resonance-fluorescence spectrum. First, in the near-noiseless limit (\(D \simeq 0\)), the peak heights increase with single-photon detuning \(\Delta\). Second, for nonzero fluctuation strength \(D\), the peak heights become largely independent of \(\Delta\); at fixed \(\Delta\), however, increasing \(D\) broadens the spectral lines and reduces their amplitudes. The peaks are located at \(\omega = 0, \pm R/2, \pm R\).
\begin{figure*}[t]
	\centering
	\subfloat[]{\includegraphics[width=0.45\linewidth]{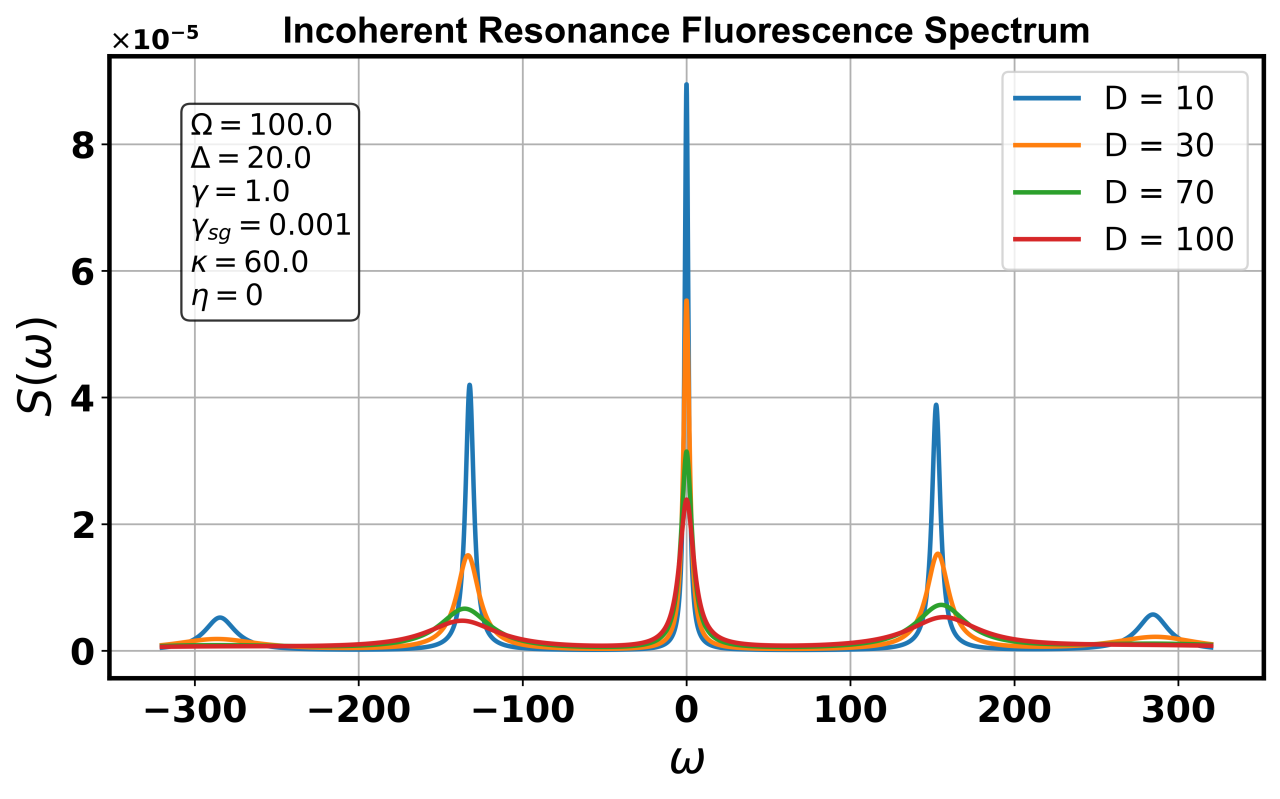}\label{resdele0R}}
	\hspace{0.01\linewidth}
	\subfloat[]{\includegraphics[width=0.45\linewidth]{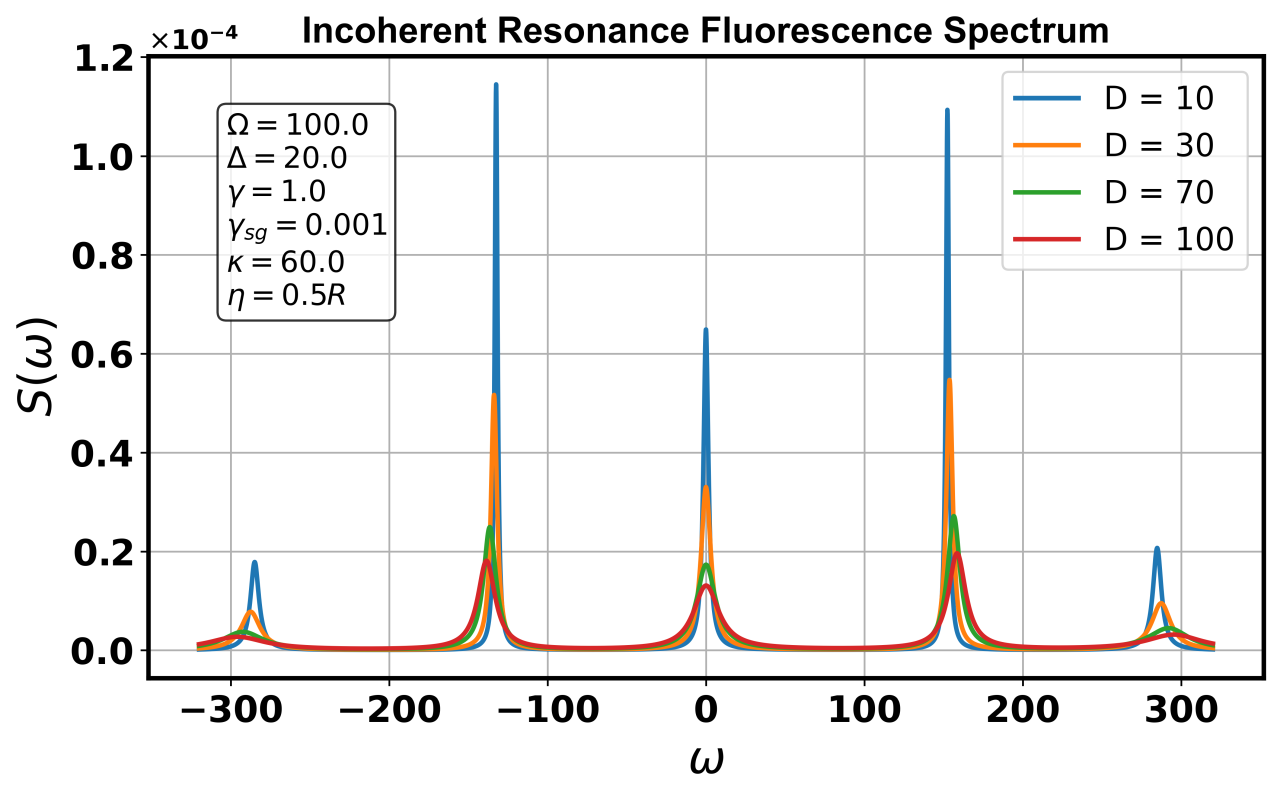}\label{resdele0.5R}} \\[5pt]
	\subfloat[]{\includegraphics[width=0.45\linewidth]{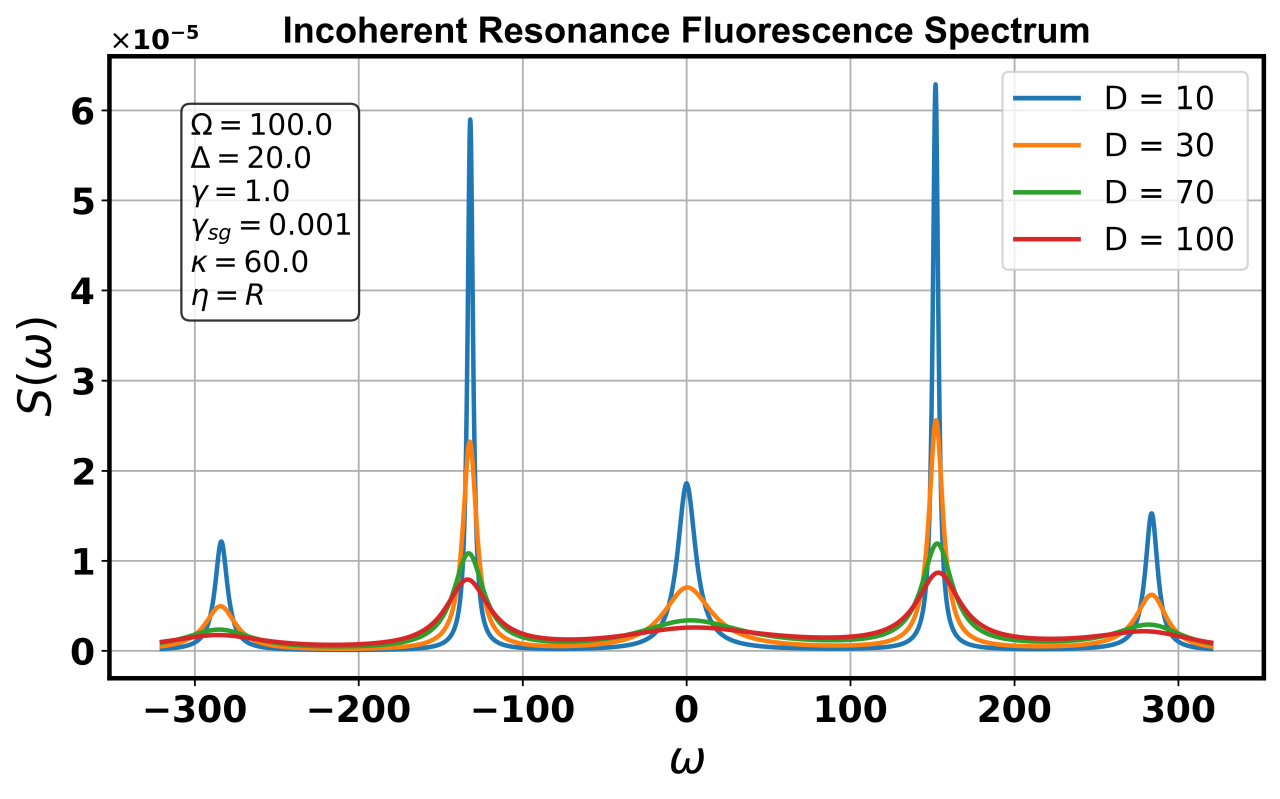}\label{resdeleR}}
	\hspace{0.01\linewidth}
	\subfloat[]{\includegraphics[width=0.45\linewidth]{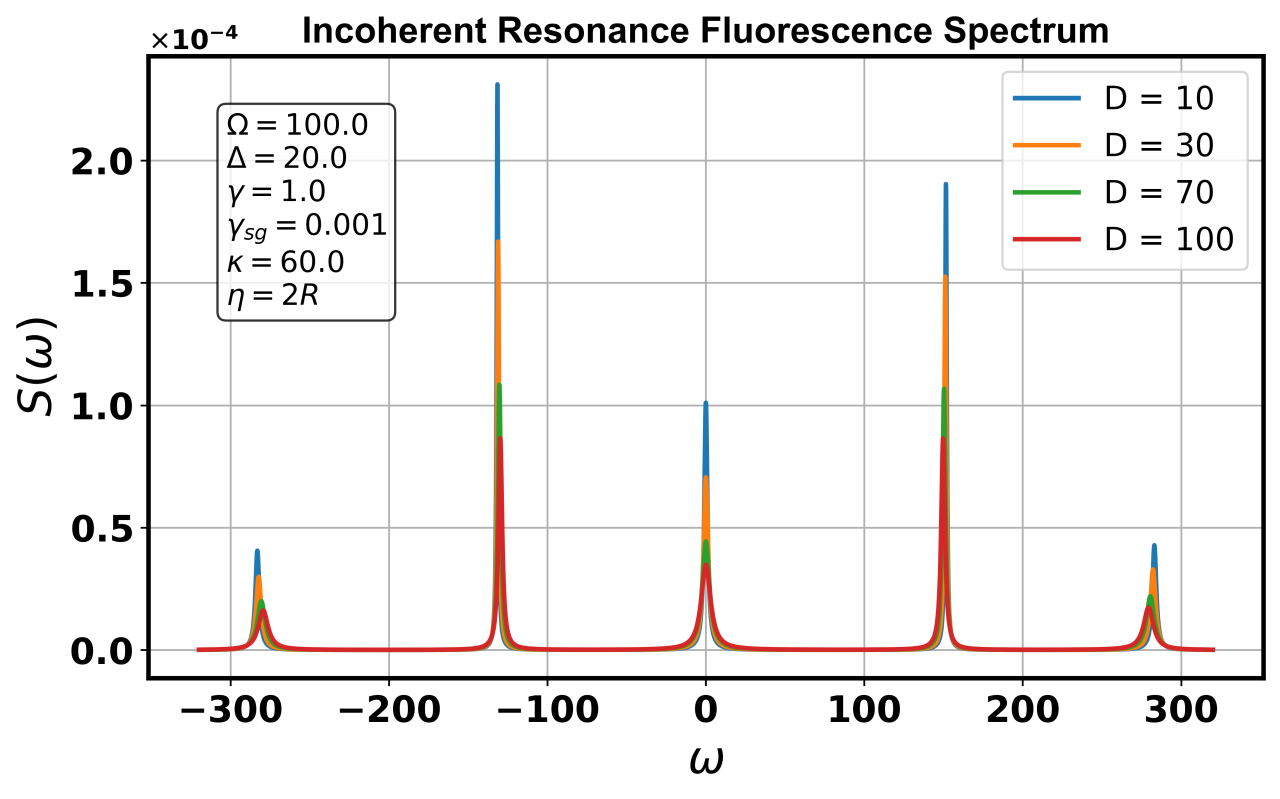}\label{resdele2R}}
	
	\caption{As in Fig.\ref{res-del}, but with $\Delta = 20$ and $\eta = 0, R/2, R, 2R$ (panels a–d). Spectra for several stochastic-field strengths $D$ are shown.}
	\label{res-del-eta}
\end{figure*}
Fig.\ref{res-del-eta} presents the incoherent resonance-fluorescence spectra across a range of stochastic-field strengths $D$ for $\eta = 0,\; R/2,\; R,\; 2R$. The spectrum for $\eta = R$ (Fig.\ref{resdeleR}) is distinct, displaying its largest peak at $\omega \simeq +R/2$ with an overall reduction in spectral intensity. This contrasts with the common pattern for $\eta = 0,\; R/2,$ and $2R$ (Figs. \ref{resdele0R}, \ref{resdele0.5R}, \ref{resdele2R}), where the dominant peak occurs at non-positive frequencies—near either $\omega \simeq 0$ or $\omega \simeq -R/2$—and grows in height with increasing $\eta$.
This reduction reflects decreased photon emission due to suppressed transition probabilities at those frequencies, indicating enhanced system stability from reduced excitation of the corresponding dressed state. As shown in Fig.\ref{eed:main}, for \(\eta = R\) the intermediate-state population \(\rho_{ee}\) remains small and nearly fixed across varying \(D\), while ground-state populations dominate. Consequently, excitation from the intermediate state is suppressed, ground-state coherence is preserved, and the system approaches partial excited-state elimination with enhanced dark-state formation, depending on the applied noise characteristics.

Across all panels, increasing the stochastic-field intensity \(D\) suppresses the spectral peaks: the central (Rayleigh) peak is most strongly reduced, while the sidebands become relatively more prominent.
\begin{figure*}[t]
	\centering
	\subfloat[]{\includegraphics[width=0.45\linewidth]{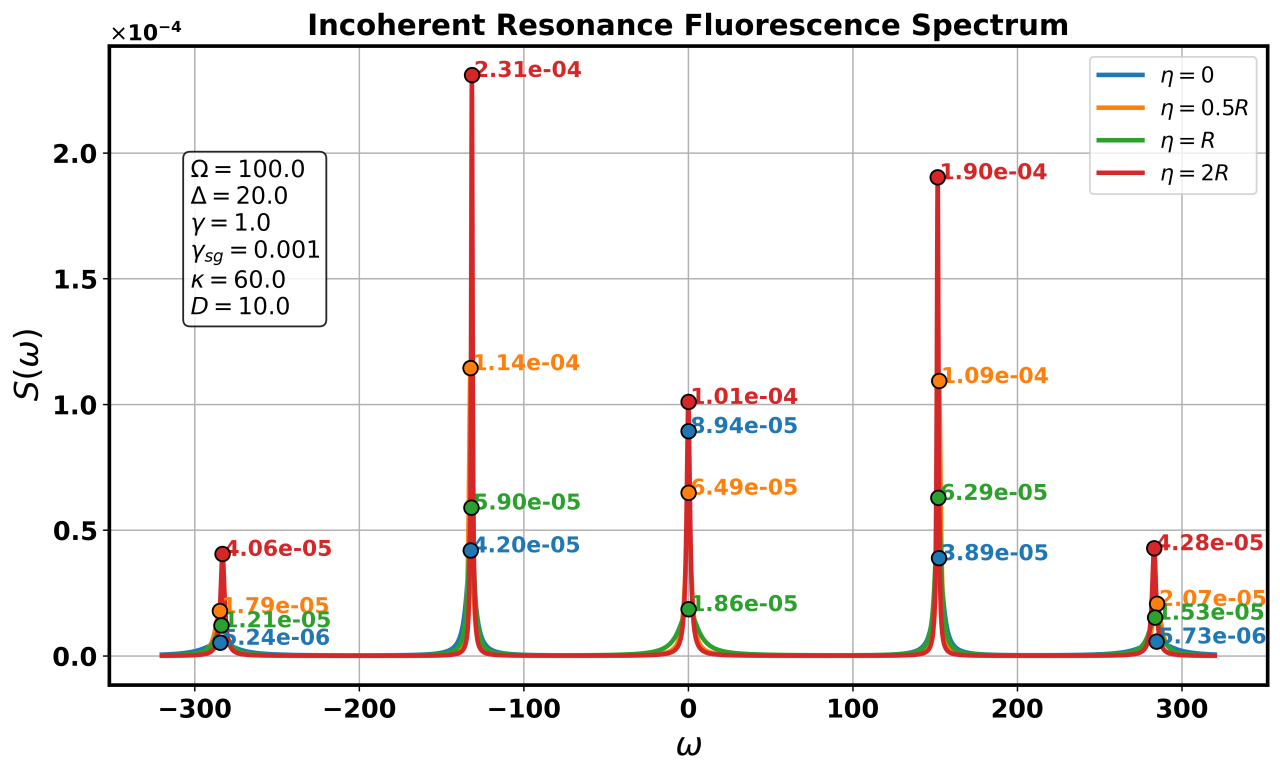}\label{rsdd10}}
	\hspace{0.01\linewidth}
	\subfloat[]{\includegraphics[width=0.45\linewidth]{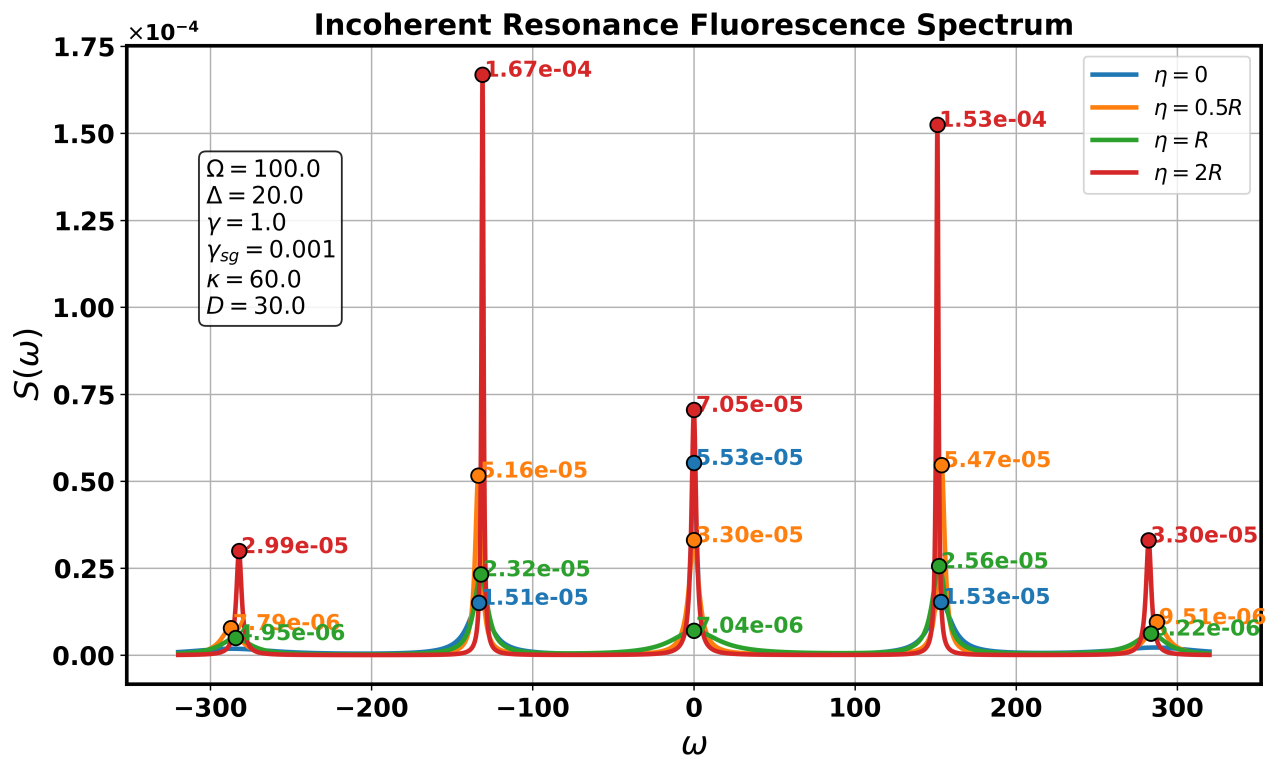}\label{rsd30}} \\
	\subfloat[]{\includegraphics[width=0.45\linewidth]{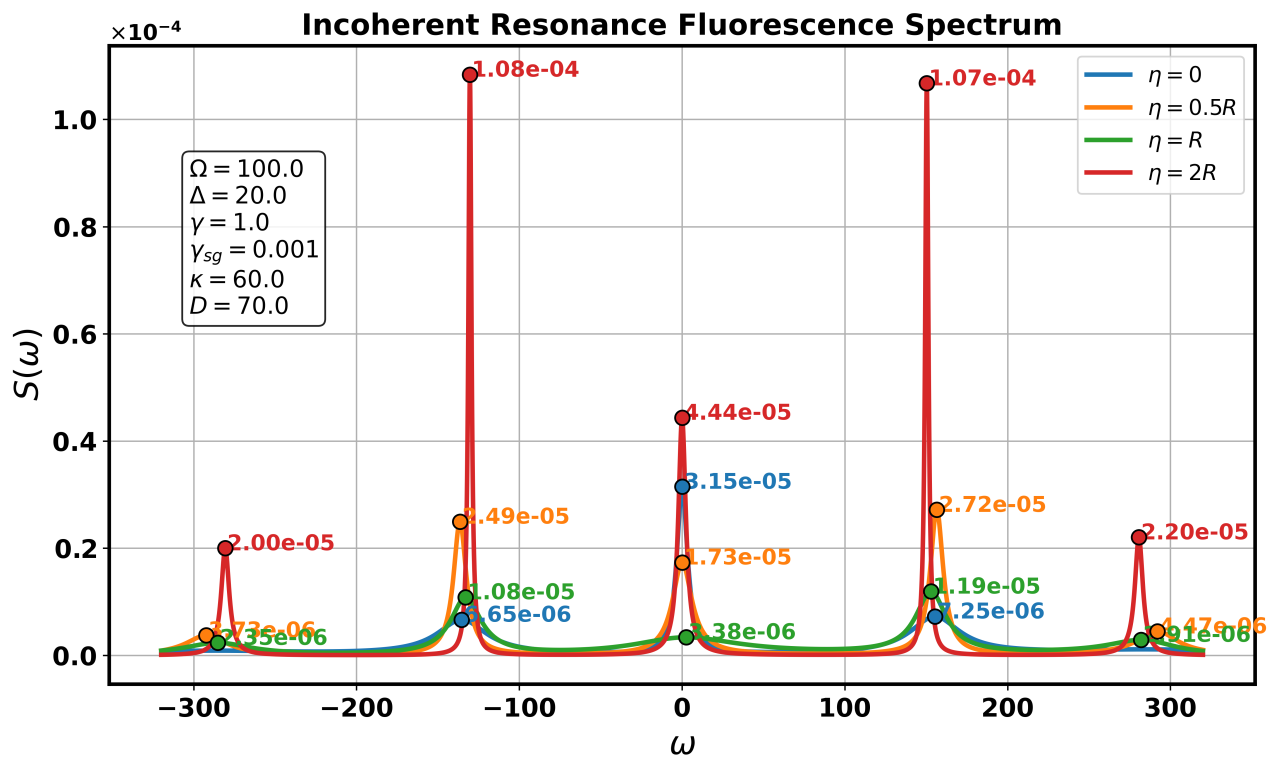}\label{rsd70}}
	\hspace{0.01\linewidth}
	\subfloat[]{\includegraphics[width=0.45\linewidth]{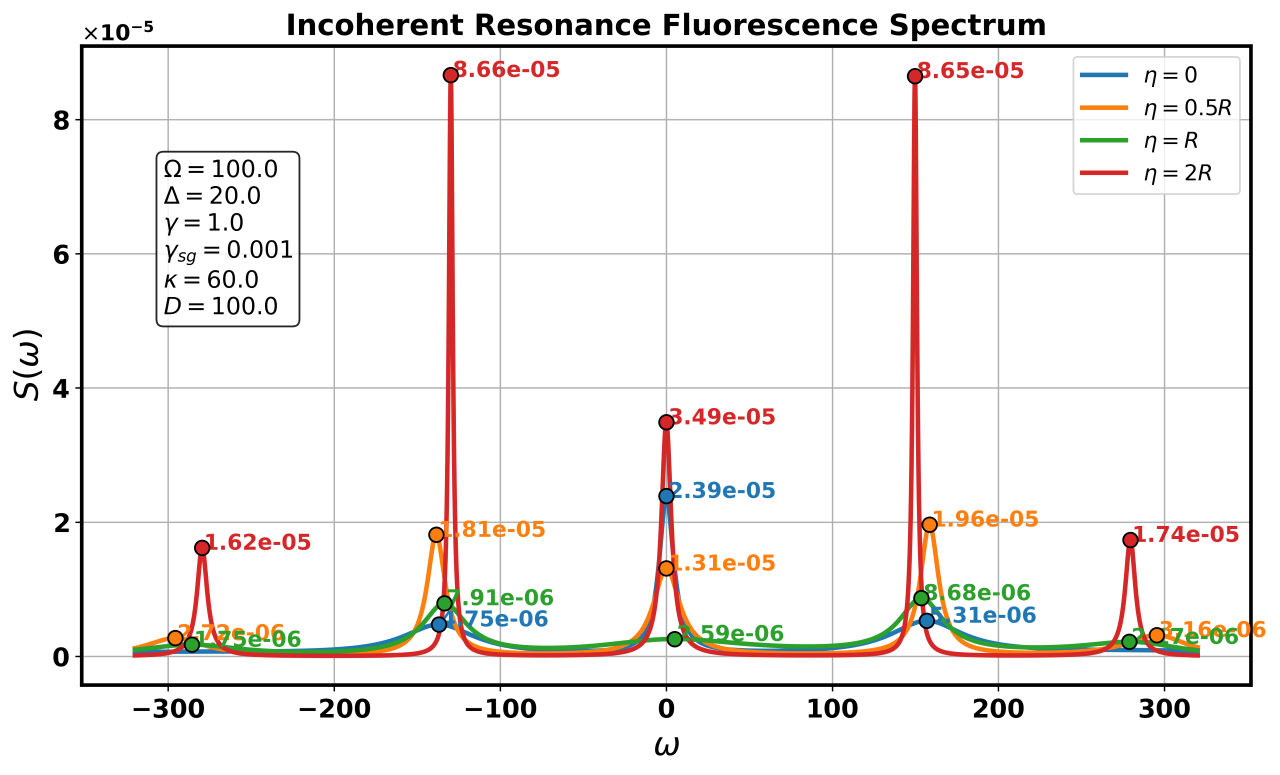}\label{rsd100}}
	
	\caption{Same parameters as Fig.\ref{res-del-eta}, but for fixed \(D=10, 30, 70, 100\) and varying \(\eta\).}
	\label{rsd:main}
\end{figure*}
Fig.\ref{rsd:main} illustrates the incoherent resonance fluorescence spectrum for $\eta = 0, R/2, R, 2R$ under increasing stochastic field strengths $D$. Evidently, higher values of $D$ result in substantial spectral suppression. Of particular note, the peak characteristics display a multifaceted dependence on $\eta$, consistent with patterns observed in Fig.\ref{res-del-eta}.\\
For the central Rayleigh-like peak, the height ordering (from greatest to least) is $\eta = 2R > 0 > R/2 > R$. In contrast, the sidebands exhibit a distinct ordering: $\eta = 2R > R/2 > R > 0$.

Resonances arise when $\eta$ coincides with dressed-state transition frequencies (e.g., $\eta = 0,  \pm R/2,  \pm R$), emphasizing the role of the stochastic field as a tunable control mechanism. In such cases, it acts as an external pump, selectively amplifying (or attenuating) dressed-state transitions. This effect is observed through variations in peak heights, offering direct insight into the atom--field dressed-state structure.

Analysis of the numerical results requires clear physical insight into the origins of the spectral peaks, their frequencies, the lineshapes, and the roles of the noise parameters. Under strong coherent driving, the atom-field interaction is most naturally described in the dressed-state basis. \\
Peaks in the incoherent fluorescence spectrum correspond to transitions between dressed states; the frequency of each peak, $\omega$, equals the energy difference between the relevant dressed states. Depending on the drive-field strength \(\Omega\), these transitions produce multi-peaked structures (typically three- or five-peaked) centered approximately at \(\omega \simeq 0, \pm R/2, \pm R\). Increasing the Rabi frequency \(\Omega\) enlarges the splitting \(R\) and shifts the sidebands farther from the center, while a nonzero single-photon detuning \(\Delta\) disrupts the spectral symmetry.

The stochastic field is modeled as a Gaussian-Markov process entering the dynamics via $Z_{\pm}$ (see Eq.\eqref{x50}). The real parts of $Z_{\pm}$ increase effective decay rates and broaden spectral lines. When the noise central frequency is resonant with a dressed-state transition (i.e., $\eta\pm nR\simeq 0, \ \ n=0,\pm 0.5, 1, 2$), its influence on that transition is amplified. This resonance enables selective control of peak intensities and dressed-state populations: by tuning the noise parameters $D$, $\kappa$, and $\eta$, one can suppress or enhance interference phenomena.
\section{Conclusion\label{sec:conclusion}}
In this work, we have analyzed the incoherent resonance-fluorescence spectrum of a three-level $\Lambda$-type atom driven by a coherent field and coupled to a classical stochastic field. Our results demonstrate that the stochastic field---typically regarded as a source of decoherence---can play a constructive role in controlling atomic emission.\\
Specifically, noise in this system serves as a resource for spectral control. For example, CPT manifests as emission suppression (appearing as a dip or a strongly weakened central peak). Appropriately tuned noise can destroy the coherence responsible for CPT, releasing the atom from the dark state and restoring the central peak. Consequently, changes in peak heights provide a quantitative probe of the robustness of interference phenomena, such as CPT, in the presence of noise. These conclusions align with our numerical results, which reveal the complex dependence of peak heights and widths on noise parameters.\\
Tuning the noise strength $D$ and the detuning $\eta$ between the noise and coherent field enables selective manipulation of dressed-state populations and transition rates. Increasing $D$ reduces peak amplitudes while preserving the overall spectral structure, allowing targeted suppression or enhancement of specific emission lines; in particular, increased noise can stabilize dark-state trapping and reduce photon emission via CPT.\\
Both amplitude- and phase-fluctuating noise thus constitute practical control handles for system modulation. This insight is relevant to quantum technologies, where realistic laser sources exhibit residual fluctuations that, though often negligible in classical contexts, can profoundly influence quantum dynamics. Accurate modeling and deliberate tuning of these fluctuations can enhance control fidelity in quantum experiments.

Overall, these findings highlight the potential of stochastic noise as a tunable parameter for manipulating atomic emission and improving spectral stability in multilevel quantum systems.
\bibliography{MyReferences}
\end{document}